\RequirePackage{snapshot}
\documentclass[a4paper,12pt]{article}
\usepackage[scale={0.8,0.9},centering,includeheadfoot]{geometry}
\usepackage{authblk}

\author[1]{Haakon Bakka}
\author[2]{Jarno Vanhatalo}
\author[3]{Janine B Illian}
\author[4]{Daniel Simpson}
\author[5]{H{\aa}vard Rue}
\affil[1,5]{ Statistics Program, CEMSE Division, King Abdullah University of Science and Technology, Thuwal 23955, Saudi Arabia}
\affil[2]{Department of Mathematics and Statistics, Faculty of Science, and
	Organismal and Evolutionary Biology Research Programme, Faculty of Bio- and Environmental Sciences,
	University of Helsinki, Gustaf H\"{a}str\"{o}min katu 2b,
	P.O. Box 68, FIN-00014 University of Helsinki, Finland}
\affil[3]{Centre for Research into Ecological and Environmental Modelling, School of Mathematics and Statistics, 
	University of St Andrews, The Observatory, Buchanan Gardens, St Andrews, Fife, KY16 9LZ, Scotland, UK}
\affil[4]{Department of Statistical Sciences, University of Toronto,
	100 St. George Street, Toronto, Ontario, Canada, M5S 3G3.}

\usepackage{natbib}
\usepackage{amsthm} 
\usepackage{amsfonts} 
\usepackage{amscd} 
\usepackage{amsmath} 
\usepackage{graphicx}
\usepackage{epstopdf}
\usepackage{pgf}
\usepackage{array, multirow}
\usepackage{caption}
\usepackage{subcaption}

\usepackage{hyperref}

\theoremstyle{definition}
\newtheorem{theorem}{Theorem} 

\newcommand{\m}[1]{\ensuremath{\mathcal{#1}}}
\newcommand{\mb}[1]{\ensuremath{\mathbb{#1}}}

\renewcommand{\d}{\ \mathrm{d} }
\newcommand{\<}{\langle}
\renewcommand{\>}{\rangle}

\newcommand{\up}[1]{\text{#1}}

\newcommand{\Cov}{\text{Cov}}

\newcommand{\object}{\texttt}
\newcommand{\package}[1]{\textbf{\textsf{#1}}}

\graphicspath{{ext/fig/}}

\newcommand{\dptavg}{AverageDepth } 
\newcommand{\dist}{Dist30m } 

\newcommand{\joetdsumsq}{RiverFlow } 
\newcommand{\joetdsumsqNospace}{RiverFlow} 
\newcommand{\swmlog}{Openness } 
\newcommand{\temjul}{TempSum } 

\title{Non-stationary Gaussian models with physical barriers}
\date{\today}

\begin{document}
	
	\maketitle

\begin{abstract}
	The classical tools in spatial statistics are stationary models, like the Mat\' ern field. 
	However, in some applications there are boundaries, holes, or physical barriers in the study area, e.g.\ a coastline, and stationary models will inappropriately smooth over these features, requiring the use of a non-stationary model.
	
	We propose a new model, the Barrier model, which is different from the established methods as it is not based on the shortest distance around the physical barrier, nor on boundary conditions.
	The Barrier model is based on viewing the Mat\' ern correlation, not as a correlation function on the shortest distance between two points, but as a collection of paths through a Simultaneous Autoregressive (SAR) model.
	We then manipulate these local dependencies to cut off paths that are crossing the physical barriers.
	To make the new SAR well behaved, we formulate it as a stochastic partial differential equation (SPDE) that can be discretised to represent the Gaussian field, 
	with a sparse precision matrix that is automatically positive definite.
	
	The main advantage with the Barrier model is that the computational cost is the same as for the stationary model.
	The model is easy to use, and can deal with both sparse data and very complex barriers, as shown in an application in the Finnish Archipelago Sea.
	Additionally, the Barrier model is better at reconstructing the modified Horseshoe test function than the standard models used in R-INLA.
	
	\emph{Keywords: Archipelago, Barriers, Coastline problem, INLA, Spatial statistics, SPDE, Stochastic partial differential equations}
\end{abstract}

\section{Introduction}
\subsection{Background}
Spatial Gaussian fields (SGFs) are widely used as model components when building spatial or spatio-temporal models for a variety of applications, e.g.\ in the Generalised Additive model (GAM) framework.
These spatial model components are used to model the residual spatial structure, resulting from unmeasured spatial covariates, spatial aggregation, and spatial noise.
In applications where no suitable covariates are available, the SGF and the intercept may be the only components in the model.

SGFs, also known as Gaussian fields or Gaussian random fields, are usually assumed to be stationary and isotropic.
An SGF is stationary if the model component does not change when the underlying map is moved.
Similarly, isotropy implies that the model component does not change when the map is rotated.
For convenience, we will use the word stationary to include both stationarity and isotropy, and also for discrete approximations of these models.
Using a stationary SGF implies the assumption that any non-stationarity in the data are covered by the other model components, e.g. by the spatial covariates, or that the data is stationary.
However, when there are physical barriers, or holes, in the study area, stationarity is an unrealistic assumption, as moving/rotating the map changes the locations of these features, and should change the model. In this case, the dependency between two observations should not be based on the shortest Euclidean distance between the locations, 
but should take into account the effect of physical barriers, and ``smooth around'' them.

\subsection{Motivating example}

\begin{figure}
	\centering
	\includegraphics[width=.6\linewidth]
	{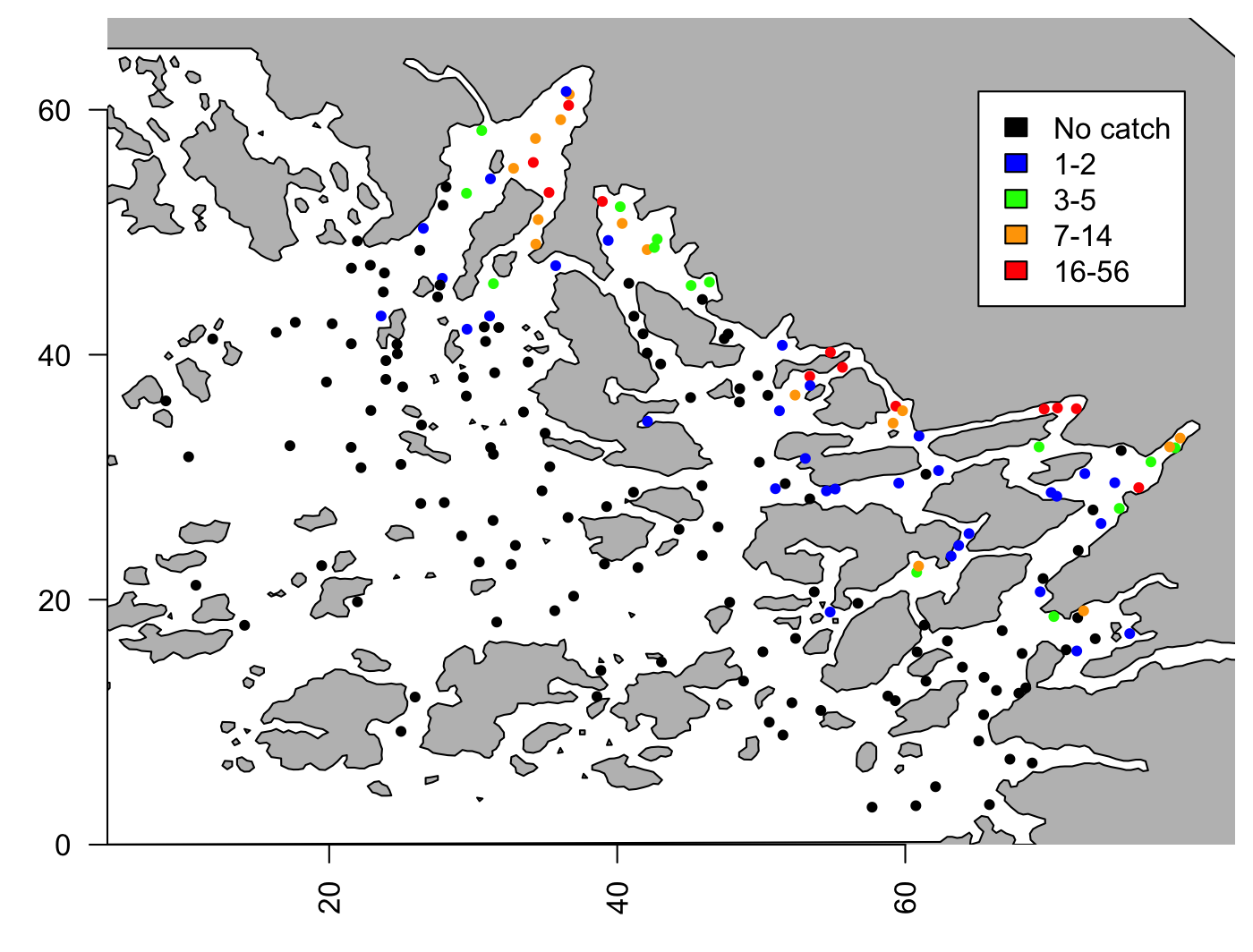}
	\includegraphics[width=.35\linewidth]
	{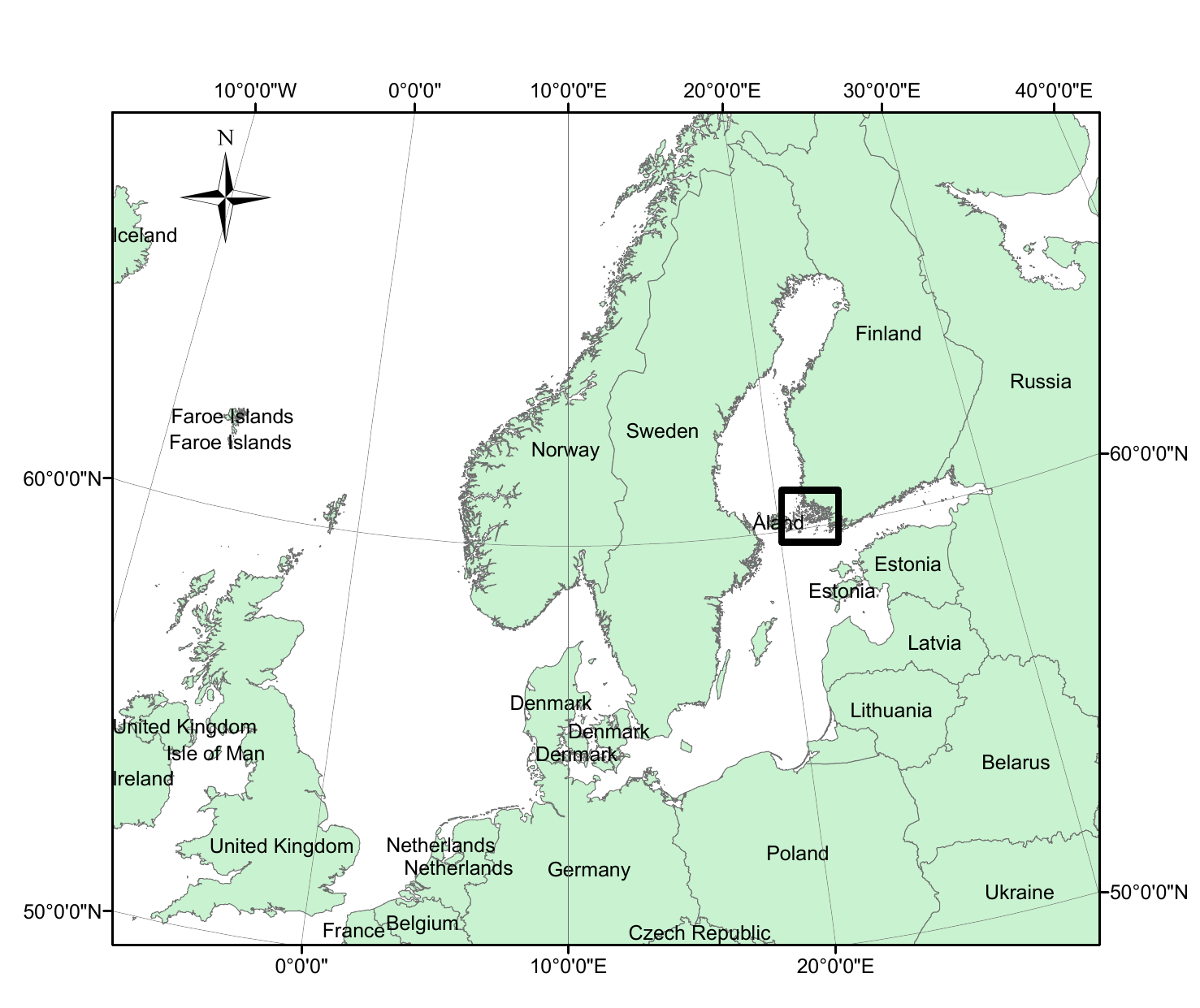}
	\caption{
		The Archipelago Sea with an example dataset on counts of smelt larvae.
		The grey region is land and the white region is water.
		This study area has many islands and peninsulas, which we do not want the spatial model component to smooth over.
		The spatial axes are in kilometers.}
	\label{fig-archip-data}
\end{figure}

The motivating example we use in this paper is an archipelago on the south-west coast of Finland, see Figure \ref{fig-archip-data}, analysed by \cite{Kallasvuo+etal:submitted} and re-analysed in Section \ref{sect-archipelago}.
This example is both complex enough to motivate a general solution and it contains many common features, including peninsulas, inlets and islands of many different sizes. 
A stationary SGF would smooth over peninsulas and islands, creating an unrealistic dependence structure for aquatic animals, and so, a non-stationary model seems more appropriate.

We will refer to the problem of physical barriers that need to be taken into account by the SGF as the \emph{coastline problem} to simplify the language, 
not implying that we only consider coastlines to be valid examples of the coastline problem.
Other examples of physical barriers include roads, power lines, mountains, and areas with different land use, and the models we discuss are able to deal with any of these.
Further, we will assume that \emph{water} is the normal area, and \emph{land} is the physical barrier; this terminology needs to be reversed when we model data on land.
Additional examples of the coastline problem can be found in the introductions by \citet{Wood2008Soapfilm} and by \citet{scott2014complex}.

\subsection{Expanding the problem formulation}
\label{sec-intro-polygon}

\begin{figure}
	\centering
	\includegraphics[height=50mm]{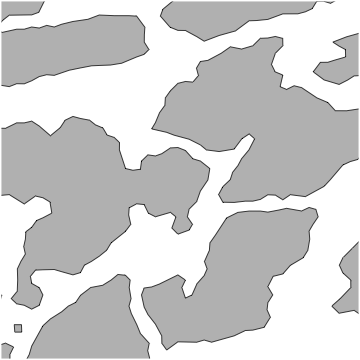} 
	\hspace{2mm}
	\includegraphics[height=50mm]{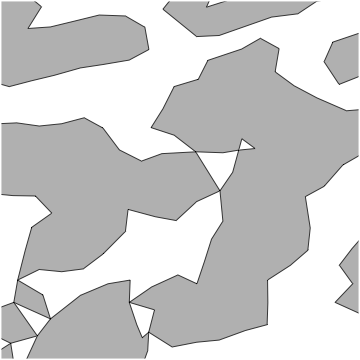} 
	\caption{Two approximations of the same coastline, at two different resolutions. Computing the shortest distances between locations in the left plot is very different from computing shortest distances in the right plot, as the East-West channel has disappeared completely. This example is a small cut-out of the study area in Figure \ref{fig-archip-data}.} 
	\label{fig-two-barriers}
\end{figure}

Another issue in the context of the coastline problem is what we refer to as the \emph{boundary polygon selection process}, explained through the following example.
A researcher desiring to model observations near the coast must represent the coastline with a polygon, as in Figure \ref{fig-archip-data}.
However, coastlines are often thought of as fractal-like, in the sense that any finite approximation will not be accurate, hence, different researchers are likely to use a different approximation for the same coastline polygons, see Figure \ref{fig-two-barriers} for an example.
Additionally, high and low tide may change the definition of the coastline.
The spatial models constructed from two polygons should differ slightly, but not dramatically.
If the same model, with a slightly different coastline approximation, 
results in different interpretations of results and predictions, the
model looses its scientific credibility.

\subsection{Literature review}
\label{sect-intro-review}

There have been several approaches to the coastline problem that focus on computing the shortest distance in water; 
\citet{wang2007low} develop the GLTPS from the Thin Plate Spline, 
\citet{scott2014complex} develop the CReSS as an improvement to the GLTPS,
and 
\citet{miller2014finite} embed the data in a higher dimensional space where the new Euclidean distances are close to the shortest distance in water, creating the MDSDS model.
A strength of the GLTPS and CReSS is that they have a flexibility parameter determining the distance at which information is assumed to have a spatial dependency, which also determines the distance at which observations no longer noticeably influence predictions through the SGF.
One weakness of these approaches is that they are significantly more cumbersome and time consuming to use than their stationary alternatives.
Furthermore, the concept ``shortest distance in water'' is not a robust concept; this distance would change abruptly between the plots in Figure \ref{fig-two-barriers}.
In this paper we do not consider any of these to be valid solutions to the stated problem, as they are not
robust to the boundary polygon selection process.

Three other approaches are based on defining boundary conditions.
\citet{Ramsay2002Spline} develop the FELSPLINE, which uses a smoothing penalty together with  Neumann boundary condition (height-curves are orthogonal to the boundary). 
One strength of this approach lies in the use of Finite Element Method (FEM) which gives a good approximation of the smoothness penalty for irregular observations.
The default SGF by \citet{Lindgren2011}, assuming the mesh is only constructed in water, also uses Neumann boundary conditions and FEM.
A weakness with these approaches, as discussed in \citet{Wood2008Soapfilm} and in Section \ref{sect-horseshoe} below is that the Neumann boundary condition is often unrealistic and severely impacts on the results.
\citet{Wood2008Soapfilm} develop the Soap-film smoother, and \citet{sangalli2013spatial} the SSR model, both models enabling the use of the Dirichlet boundary condition (a known value/function along the boundary).
The main strength with the Dirichlet boundary condition is that when the true boundary values are known, not only will this give the correct result at the boundary, but the spatial field will smooth these values far into the water.
As this condition acts like ``perfect observations'', observations near the boundary are not necessary for estimating the SGF near the boundary, and the result will have very narrow estimation intervals near the boundary.
This strength is also a weakness: If the employed boundary condition is slightly wrong, the true intensity near the boundary is not covered by any estimation intervals,
and in practice it is unlikely that the true boundary values are known explicitly.
Indeed, in our guiding example (Figure \ref{fig-archip-data}), we cannot use a Dirichlet condition of ``no fish at the boundary'' as this would imply that there are almost no fish close to the boundary either, which is clearly wrong.
In the rare case when there is sufficient certainty about true value of the SGF at the boundary, as well as the propagation through the correlation structure of the SGF, we recommend using the Dirichlet condition, but for the rest of this paper we assume that the boundary values are unknown.

One way to remedy the weakness of the Dirichlet boundary condition is to put a separate model on the boundary, and then model the SGF conditionally on this boundary model, as \citet{Wood2008Soapfilm} do for the Soapfilm smoother.
A strength of this approach compared to the simple Dirichlet approach is that it can reflect large uncertainties near the boundary.
The model for the boundary, however, is very influential on the results near the boundary, and 
it is not clear what would constitute a good model for that role, nor how the model should change when the boundary is deformed.
The main weakness of this approach is that it can fail to produce reliable models in cases with sparse data and complex boundaries.
In our motivating example there are more islands than observation locations, which leads to a massive overparametrisation when creating a boundary model for each island.

\subsection{Requirements for a new solution}
Here, we set out five desirable properties of a solution to the coastline problem.
These properties are inspired by the literature review and formulated to ensure that the main strengths of previous approaches are maintained, while the main weaknesses are overcome. 
We readily acknowledge that different scientists have different requirements for their models; 
we do not wish to imply that this list is the only possible list of requirements, 
but we recommend any reader who agree with this list to use the model we present in this paper.
The SGF should
\begin{description}
	\item[Property 1] be robust to the boundary polygon selection process,
	\item[Property 2] have a computational cost close to the stationary alternative,
	\item[Property 3] not be much more difficult for a researcher to use in practice, compared to the stationary GRF,
	\item[Property 4] include a flexibility parameter, a \emph{range} parameter, determining the distance of spatial similarity, as described below, and,
	\item[Property 5] not introduce any new assumptions that are not realistic.
\end{description}

\begin{figure}
	\centering
	\includegraphics[height=35mm]{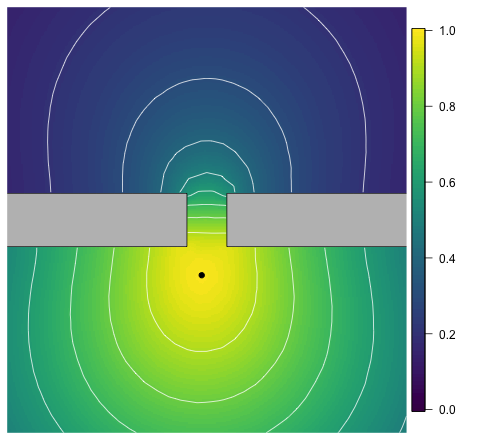} 
	\includegraphics[height=35mm]{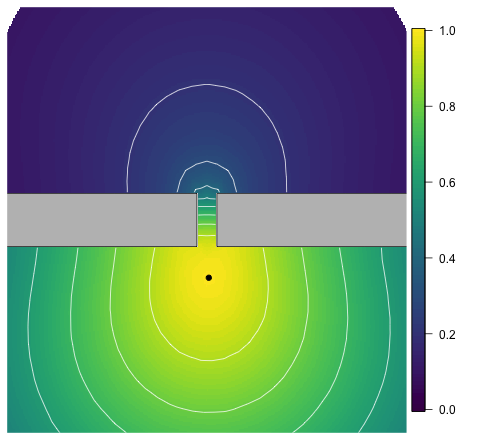} 
	\includegraphics[height=35mm]{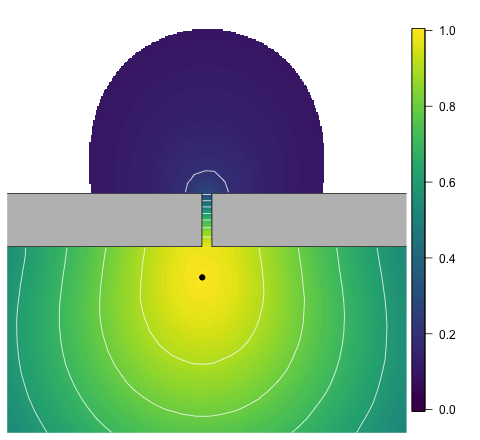} 
	\includegraphics[height=35mm]{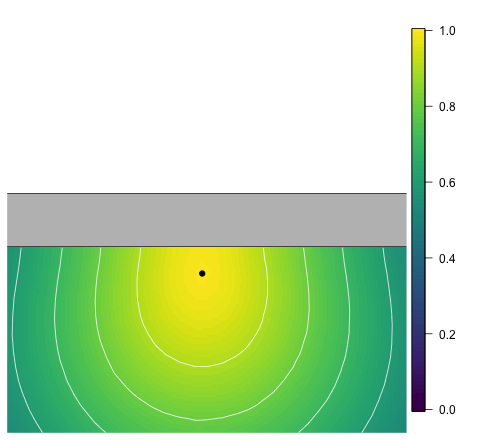} 
	\caption{Correlation plots of the SGF in \citet{Lindgren2011} where there is an opening in the land barrier. 
		All parameters in the three plots are the same, except the gap widths, which takes the values 0.4, 0.2, 0.1, and 0.
	} 
	\label{fig-neumann-robust}
\end{figure}

Property 1 was discussed in Section \ref{sec-intro-polygon}. 
In Figure \ref{fig-neumann-robust} we illustrate the behaviour of the default SGF with Neumann boundaries in \citet{Lindgren2011} when modelling a small channel in a coastline. 
As the channel becomes smaller, and is eventually closed, the dependency between the south and the north shrinks towards zero.
This illustrates that the behaviour of this  model component depends continuously on the width of the channel, hence we believe it is robust to the boundary polygon selection process.


For Property 2, the computational cost is a clear bottleneck in many applications, as researchers need to fit several models to compare predictive performance and investigate the stability of inference results.
We consider the $\m O(n^3)$ cost of computing all possible distances between $n$ datapoints, as used in e.g.\ \citet{scott2014complex}, or even the $\m O(n^2)$ cost of writing down the full spatial covariance matrix, to be too expensive for a general solution to the coastline problem.

Another important problem, although less well defined, is the cost of time for a researcher to employ the model in practice.
Property 3 is of little interest from a mathematical point of view, but to achieve widespread use this property is essential.
To reduce the effort needed, constructing the model component should be automatic, 
in the sense that there should be few, if any, 
additional specifications or computations that the user need to do,
and that the results should be numerically stable
even for complex coastlines, such as in our motivating example.
The SGF is often used as a nuisance parameter, so we should not add any new steps in model construction, compared to the stationary SGF, that need to be tuned and/or assessed by model comparison.
Models constructed from this SGF should be able to fit both sparse and rich data, without having to consider whether the dataset is informative enough to do inference with a non-stationary SGF.

Property 4 was discussed in Section \ref{sect-intro-review}.
This parameter increases flexibility in the sense that, if a user does not want to infer it from the data, it can instead be fixed to a constant; intrinsic smoothers can be recovered by letting this constant go to infinity.

Property 5 is a catch-all for avoiding new models with problematic behaviour. 
If, for example, we can solve the coastline problem by introducing an artificial partition of our dataset, or by moving the locations to arbitrary new locations, such an approach may satisfy Property 1 to 4, but should not be considered a valid approach.
When new models are developed, as in this paper, we need to shed light on the behaviour of the model to the best of our ability.
Common sense, or application experts, can then determine whether the assumptions introduced by the new approach are reasonable.
Property 5 is the only property that the default model by \cite{Lindgren2011} does not satisfy; in Section \ref{sect-horseshoe} we show that the Neumann boundary condition is unrealistic and has a strong impact.

\subsection{Outline of the paper}

The rest of the paper is organised as follows.
In Section 2 we motivate and define the Barrier model, a new approach to the coastline problem satisfying all of the stated properties.
In Section 3 we show how to construct the model component as a Gaussian precision matrix conditional on hyper-parameters.
In Section 4 we run the standard modified horseshoe reconstruction problem \citep{Wood2008Soapfilm}, to compare the new model to the Neumann model and the stationary alternative.
In Section 5 we analyse the motivating example, fish data from the Finnish Archipelago Sea.
In the supplementary material we provide more details on all the models, and tutorials for how to use the Barrier model.

\section{The Barrier SGF}
\label{sect-barrier}

\subsection{Background}

One of the most widely used models for the SGF is the 
Mat\' ern model \citep{Whittle1954}, with justification \citep[e.g.][]{diggle2010historical}, and applications varying from simple geostatistical models with Gaussian likelihoods to marked point patterns \citep{illian2012toolbox}.
The interest in applying the Mat\' ern model seems to be increasing, partly because of the INLA approach \citep{Rue2009} and the SPDE approach \citep{Lindgren2011}, as this enables fast Bayesian inference, and is relatively user friendly (see e.g.\ \citet{blangiardo2015spatial}).
A few notable examples include;
analysing transmission intensity of malaria \citep{art615},
modelling under-five and neonatal mortality \citep{golding2017mapping}, and
assessing the impact of control measures on malaria in Africa \citep{art618}.

The Mat\' ern field is a Gaussian random field, i.e.\ a continuously indexed random variable where the indices are a subset of $\mb R^n$, and any finite collection of indices gives variables that are jointly multivariate Gaussian.
We base the Barrier SGF on the Mat\' ern model, in dimension 2, and, in the absence of land, it reduces to this model.
We fix the smoothness parameter, as this parameter is in some cases not identifiable, choosing $\nu=1$ as this value is both convenient to work with and provides reasonably smooth fields.
We refer to \citet{Bolin17rational} and \citet{bolin2017numerical} for how to extend SPDE models to other values of $\nu$.

One way to represent the Mat\' ern field $u(s)$ is by giving the covariance as a function of distance between two points. 
Letting $d$ be the distance between two arbitrary points, $d=||s_i - s_j ||$, the covariance function is
\begin{align}
C(d) &= \sigma_u^2 \frac{d \sqrt{8}}{r} K_1 \left( \frac{d \sqrt{8}}{r} \right).
\label{eqMaternCovar}
\end{align}
Here, $\sigma_u $ and $r$ are constants, and $K_1$ is the modified Bessel function of the second kind. 
The subscript on $\sigma_u$ clarifies that this is the marginal standard deviation of the model component $u$.
Note that we have re-parametrised the traditional Mat\' ern covariance function with range $r = \rho/\sqrt{8}$ where $\rho$ is the traditional length scale parameter (see \citet{Lindgren2011}).
The range $r$ is interpretable because the correlation between two points that are $r$ units apart is near 0.1, and $r$ is
approximately the smallest distance from high value to low value regions of a random sample from the SGF.
If we were to follow previous approaches, and defined a shortest distance, we might try to replace the $d$ in the above equation with an approximate shortest distance around land.
This, however, would not satisfy Property 1.

\subsection{Motivating the new approach}
To motivate our approach we interpret the Mat\' ern field as a Simultaneous Autoregressive (SAR)
model on a grid giving an approximate representation (for a fixed range).
Let $U_{i, j}$ be the random variable in grid cell $[i,j]$, and let $z_{i, j}$ be iid zero mean Gaussians,

$$U_{i, j} - k\left( U_{i-1, j} + U_{i, j-1} + U_{i+1, j} + U_{i, j+1} \right)  = z_{i, j} , $$
see Equation (5) and onwards in \cite{Lindgren2011}.

When interpreting of the Mat\' ern model in this way, the relevant ``distance'' is not the shortest distance, but rather a collection of all possible paths from one location to another; and the dependency between two points relies on all the paths that exists between them.
Intuitively, what we want to do is to remove the collection of paths crossing land;
this also implies that the new distance will not be the new shortest distance, but will be an indirect result of the new collection of available paths.
For computational reasons, we will not completely remove any paths, but instead weaken the dependency along those paths to almost zero.

The SAR model is defined for the entire area, including land.
However, the SAR model over land only exists as a computational trick, and no fitting or prediction will be done on land.
To illustrate the SAR model at the boundary between water and land, consider an example with a regular grid where land is to the right of origo, filling the space vertically, so grid cell $U_{1, 0}$ is on land, while $U_{0, 0}$ is not.
The equation for the SAR model at $U_{0, 0}$ then becomes
$$(1 - k_2) U_{0, 0} - k\left( U_{-1, 0} + U_{0, -1} + (1-k_3) U_{1, 0} + U_{0, 1} \right)   = z_{0, 0},$$
where $k_3$ reduces the dependency ($k_3 < 1$), and $k_2$ can be used to make the SAR well behaved.
Selecting the appropriate values for $k_2$ and $k_3$, for all the ``SAR equations'' near the boundary, would be an almost impossible task, if it was not for a link between SAR/CAR models and Mat\' ern models through the stochastic partial differential equation (SPDE) approach \citep{Lindgren2011}, as explained below.

\subsection{The Barrier SGF}

The stationary Mat\' ern field is the (weak, stationary) solution $u(s)$ to the stochastic partial differential equation
\begin{align}
u(s) - \nabla \cdot \frac{r^2}{8} \nabla u(s) =   r \sqrt{\frac{\pi}{2}} \sigma_u \mathcal{W}(s),
\label{eqSPDEstat}
\end{align}
where $u(s), s\in \Omega \subseteq \mathbb{R}^2$ is the Gaussian field, $r$ and $\sigma_u$ the same constants as in equation \eqref{eqMaternCovar}, $\nabla = \left(\frac{\partial}{\partial x}, \frac{\partial}{\partial y} \right)$, and $\mathcal{W}(s)$ denotes white noise. 
For further details, see \citet{Lindgren2011}; we have re-parametrised their equation (2) and fixed $\alpha=2$.
Details on the interpretation of this SPDE and how to solve it through the FEM can be found in \citet{bakka2018solve}.

On land we introduce a different Mat\' ern field, with the same $\sigma$ but a range close to zero, to remove the correlation there. 
The Barrier SGF 
$u(s)$ is the solution to 
\begin{align}
u(s) - \nabla \cdot \frac{r^2}{8} \nabla u(s) &=   r \sqrt{\frac{\pi}{2}} \sigma_u \mathcal{W}(s), \up{ for } s \in \Omega_n \nonumber \\
u(s) - \nabla \cdot \frac{r_b^2}{8} \nabla u(s) &=   r_b \sqrt{\frac{\pi}{2}} \sigma_u \mathcal{W}(s), \up{ for } s \in \Omega_b,
\label{eq-barr-spde}
\end{align}
where $\Omega_n$ is the normal area, $\Omega_b$ is land, i.e.\ the physical barrier, and their disjoint union gives the whole study area $\Omega$. 
To achieve Property 2, we do not include additional parameters that require tuning, inference, or model assessment, and hence the parameter $r_b$ is taken to be a fixed fraction of the range $r$, e.g. $r_b= r/10$.  
We choose the solution $u(s)$ that is continuous and satisfies the so-called ``natural matching conditions'' (see \citet{gander2015optimized} Section 5), 
$$\frac{r^ 2}{8} \frac{\partial u_n}{\partial n_b} = \frac{r_b^ 2}{8} \frac{\partial u_b}{\partial n_b}, $$
where $\frac{\partial u_n}{\partial n_b} $ is the partial derivative orthogonal to the boundary  (i.e.\ in direction $n_b$) just inside the normal area, and $\frac{\partial u_b}{\partial n_b}$ is the same derivative just inside the land area.

To prove existence and uniqueness, we view the SPDE through operators on Hilbert spaces.
Let $\Omega \subset \mb R^2$ be a polygonal domain, and
$H$ be the subspace of $L^2(\Omega )$ where
all functions satisfy Neumann boundary conditions on $\Omega $.
The operator 
$$ L = \kappa(s)  + \nabla \cdot a(s) \nabla, $$
with $\kappa, a \in L^\infty$,
$\kappa(s) >\kappa_0 >0$ and $a(s) > a_0 > 0$,
is defined on $\dot H^2 = \mathcal D(L)$ which is dense in $H$.
\begin{theorem}
  \label{theorem-existance}
The equation
$$ L u = \m W,$$
has a unique solution $u \in L_2(\Pi; H)$, $\Pi$-a.s.
\end{theorem}
For proof, see appendix \ref{app-existence-proof}.

\section{Finite dimensional representation of $u(s)$}
\label{sect-barr-finite}

In this section we detail how to represent the continuous GF $u(s)$ with a finite dimensional approximation, 
and how to compute the sparse precision matrix $Q$ for the coefficients of this approximation.
We use a linear finite element approach that approximates the solution with a (continuous) piecewise linear function, similar to \citet{Lindgren2011}.
This is to avoid any extra interpolation approximations after the solution is computed, which would be the case with a grid based approach.

We rewrite equation \eqref{eq-barr-spde} to 
\begin{align*}
\left[ 1 - \nabla \frac{r(s)^2}{8} \nabla \right] u(s) &= r(s) \sqrt{\frac{\pi}{2}}   \m W(s) \\
r(s) &= r_q  \text{ on } \Omega_q,
\end{align*}
where the domain $\Omega $ is a disjoint union of $\Omega_q$ for $q=1,2,...,k$, and with Neumann boundary condition on $\partial \Omega$. 
This is a minor generalisation; we use $k=2$ in the Barrier SGF.

Irregular outer boundaries are known to cause numerical artefacts and unrealistic behaviour, to avoid this issue we extend the study area and make it convex. 
This moves the boundaries away from the data, avoiding an impact of the boundary condition on the fitted model (see \citet{Lindgren2011} appendix A.4).
Having a regular outer boundary is a common assumption for many numerical techniques and approximations  \citep{grisvard1985elliptic}, and inference can later be restricted back to the study area.

\begin{figure}
	\centering
	
	\includegraphics[width=0.33\linewidth]
	{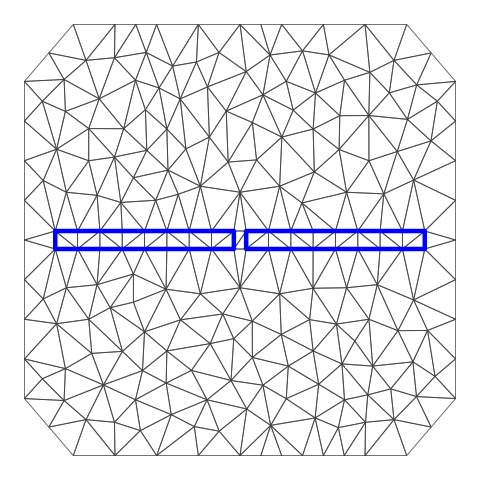}
	\includegraphics[width=0.33\linewidth]
	{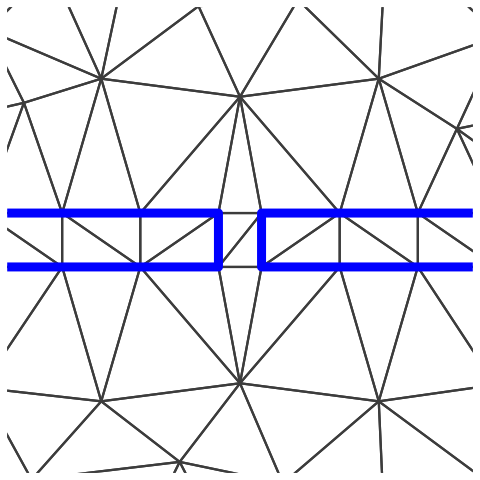}
	\caption{Example mesh  for a coarse version of the example in figures \ref{fig-neumann-robust} and \ref{fig-barrier-gaps}.
		The mesh has been extended to move the outer boundary far away from the region of interest, and the blue rectangles denote the land area.
		The second plot is a zoom-in of the first.}

	\label{fig-q-image}
\end{figure}

After a mesh has been selected, as in Figure \ref{fig-q-image}, the linear finite elements $\psi_i(s)$ are defined to be piecewise linear on this mesh, taking the value 1 in node $i$, and the value 0 in all other nodes.
The spatial field approximation $\tilde u$ is then
$
\tilde u(s) = \sum_{i=1}^n u_i \psi_i(s),
$
where $n$ is the number of basis functions (one for each mesh node), and $\tilde u_i$ are Gaussian random variables, with precision matrix $Q$.

When solving this SPDE with finite elements, the equation is re-interpreted in the following weak form,
\begin{align}
\left\langle \psi_j(\cdot ), \left[ 1 - \nabla \frac{r(\cdot)^2}{8} \nabla \right] \tilde u(\cdot)\right\rangle
&= \left\langle\psi_j(\cdot), r(\cdot) \sqrt{\frac{\pi}{2}}   \m W(\cdot)\right\rangle, \label{eq-weakform}
\end{align}
meaning that the joint distribution, over $j$, of the left hand side is equal in distribution to the joint distribution on the right hand side.
The inner product
$\< f, g \> = \int f(s) g(s) \d s. $

Define the matrices
\begin{align}
J_{i,j} &= \<\psi_i, \psi_j \> = \int \psi_i(s) \psi_j(s) \d s \\
 \label{eq-dq} (D_q)_{i,j} &=  \<1_{\Omega_q}\nabla \psi_i , \nabla \psi_j \> = \int_{\Omega_q} \nabla \psi_i (s) \nabla \psi_j(s) \d s\\
(\tilde C_{q})_{i,i} &= \<1_{\Omega_q} \psi_i, 1 \> = \int_{\Omega_q} \psi_i(s) \d s 
\end{align}
which are the basic ingredients in the finite element method. 
Writing $\tilde u(s)$ as a linear combination of elements in equation \eqref{eq-weakform}, and putting the resulting coefficients into matrix form, we get
$A \tilde u =  \epsilon $
where
\begin{align}
A &= J-\frac{1}{8}\sum_{q=1}^k r_q^2 D_q 
\end{align}
and $ \epsilon$ is multivariate Gaussian with 
\begin{align}
\Cov( \epsilon) &\approx \tilde C = \frac{\pi}{2} \sum_{q=1}^k r_q^{2} \tilde C_q,
\end{align}
which is diagonal, making the $\epsilon_i$ independent.
To get the approximation $\tilde C$ we follow the arguments of \cite{Lindgren2011}.
Since $A$ is symmetric, 
$
Q = A \tilde C^{-1} A.
$
For computational efficiency it is important to store the matrices $J, C_q$, and $D_q$ so that $Q$ can be computed quickly for new values of $(r_q)_q$.

\begin{figure}
	\centering
	
	\includegraphics[width=0.45\linewidth]
	{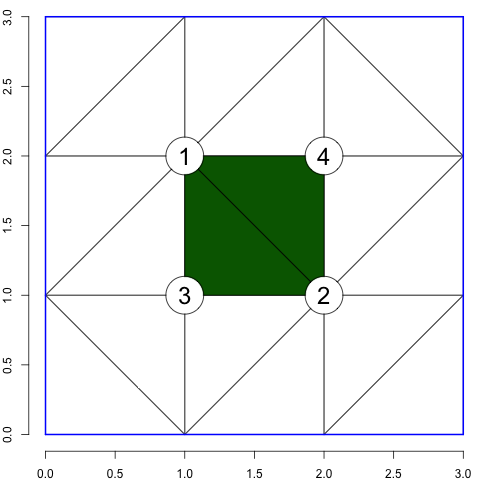}

	\caption{Example simple mesh used for computing example $D_q$ matrices in Section \ref{sect-barr-finite} .
}
	\label{fig-dq-matrices-square}
\end{figure}

The difference between a stationary SGF and the Barrier SGF can be understood from the $D_q$'s, see Figure \ref{fig-dq-matrices-square} for an illustration.
In this example, 
$$ 	D_1 = 
\begin{bmatrix}
3.0  & .    & -0.5 & -0.5 \\
.    & 3.0  & -0.5 & -0.5 \\
-0.5 & -0.5 & 3.0  & .    \\
-0.5 & -0.5 & .    & 3.0
\end{bmatrix} ,	
\qquad
D_2 = 
\begin{bmatrix}
1.0  & 0.0  & -0.5 & -0.5 \\
0.0  & 1.0  & -0.5 & -0.5 \\
-0.5 & -0.5 & 1.0  & .    \\
-0.5 & -0.5 & .    & 1.0
\end{bmatrix},
$$
with $q=1$ representing the white normal area, and $q=2$ representing the green land area in the figure.
The numbers in circles show the central location of each finite element, which corresponds to rows and columns in the matrices.
The connection between element 1 and 2 is structurally zero in $D_1$ signified by a ``.'', since no white triangles are connecting node 1 and 2, while it is numerically zero in $D_2$ as the inner product of the derivatives is zero (equation \eqref{eq-dq}).
Elements 3 and 4 share no triangles and so has structural zeroes in both matrices.
For nodes 3 and 4, 1/4 of the area of neighbouring triangles are green and the other 3/4 are white, and the triangles are symmetric around the node explaining the entries on the diagonals.
The diagonal entries for node 1 and 2 is slightly more complicated, but can be found by computing the integrals.

\begin{figure}
	\centering
	\includegraphics[height=35mm]{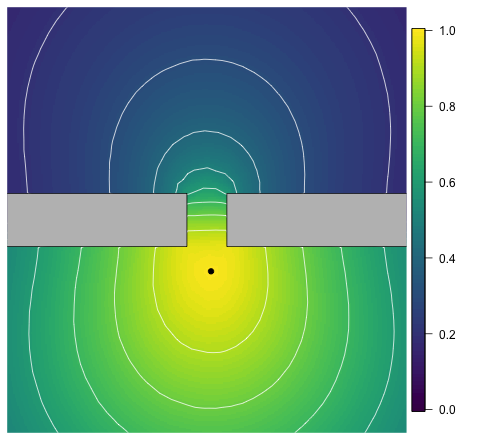} 
	\includegraphics[height=35mm]{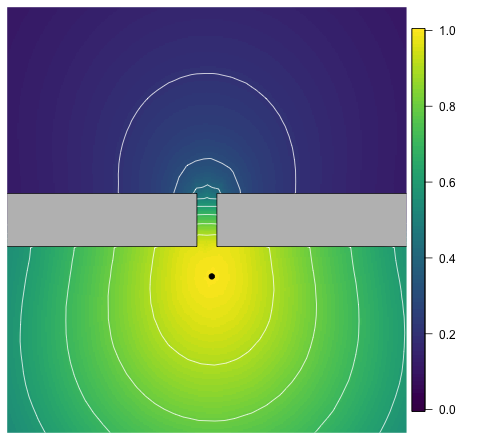} 
	\includegraphics[height=35mm]{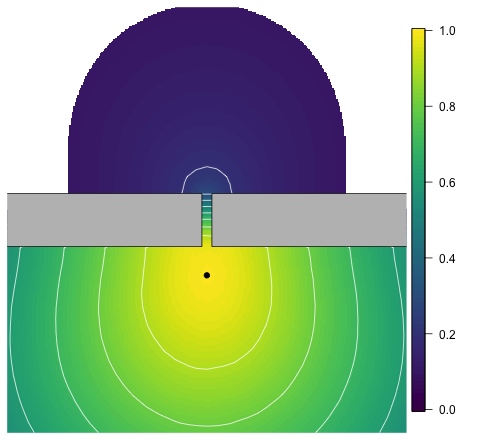} 
	\includegraphics[height=35mm]{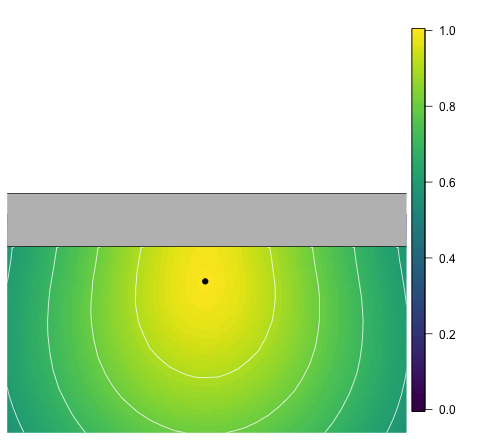} 
	\caption{Correlation plots of the same scenarios as in Figure \ref{fig-neumann-robust}, showing that the Barrier SGF is also robust to the boundary polygon selection process.}
	\label{fig-barrier-gaps}
\end{figure}

In Figure \ref{fig-barrier-gaps} we show that the new SGF is robust to the boundary polygon selection process, satisfying Property 1.
Figure \ref{fig-q-image} shows that the precision matrix is very sparse.
The computationally demanding operation when using the Barrier SGF is a repeated Cholesky factorisation, which is a factorisation of a matrix with the same sparsity structure as for the stationary SGF, resulting in the same computational cost, $\m O(n^{3/2})$ where $n$ is the number of nodes in the finite element mesh \citep{Lindgren2011}, 
hence the Barrier SGF
satisfies Property 2.

The only step that needs to be assessed by the user is the step of mesh construction, including whether the boundaries have a reasonable representation in the mesh, 
but this step is also needed when using the stationary SGF.

For the marginal standard deviation of the field at a location $s$ we have two options.
One is to use the spatially varying standard deviation we get from solving the above equations, 
which results in higher prior standard deviation in inlets.
The second option is to rescale the precision matrix so that all the marginal standard deviations are 1.
This can be done by finding the maginal variances (finding the inverse diagonal of the precision matrix through sparse computations), and rescaling the precision matrix using these values.
In this paper, we have chosen the first option mainly for its interpretation; we believe that this non-constant marginal spatial uncertainty is \emph{a priori} reasonable.
The SGF is a solution to an SPDE and represents the ``average location'' of a randomly moving individual in the physical space defined by the differential equation. 
E.g. in the archipelago example, narrow inlets are places 
that an individual is less likely to visit as opposed to the large open sea areas but if the individual gets into an inlet, it will remain in there for a longer period.
This leads to higher standard deviations compared to a stationary SGF; e.g.\ either there is a lot of fish, or there are almost no fish, in the inlet.

\section{The modified Horseshoe reconstruction problem}
\label{sect-horseshoe}

\begin{figure}
	\centering
	\includegraphics[width=.6\linewidth]
	{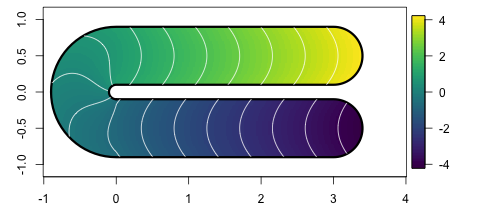}
	\caption{The true function that is to be reconstructed in the modified horseshoe example.
		The white lines are height curves at regular intervals.
	}
	\label{fig-horseshoe-trueonly}
\end{figure}

In this section we illustrate the difference between the stationarity assumption, the Neumann assumption, and the new model, through reconstructing a known test surface, and we study whether the gain in reconstruction quality when using the new model is substantial or minor.
The original Horseshoe test surface was developed by \citet{Ramsay2002Spline}, but happens to fulfil the Neumann boundary condition, and so, any approach satisfying the Neumann boundary condition has unfairly good performance.
To fix this, \citet{Wood2008Soapfilm} modified the test function so that it did not have any of the typical boundary conditions (Dirichlet or Neumann), 
see Figure \ref{fig-horseshoe-trueonly} for this test function.

We compare reconstructions from three different models, all of them hierarchical Bayesian models with Gaussian observation noise, fitted with \texttt{R-INLA} \citep{Rue2017bayesian}.
In this paper we use the term ``model'' for the entire prior model, including observation likelihood, and the term SGF for the spatial component in the model, indexed by two hyper-parameters $r$ and $\sigma$. 
MB is a model using the Barrier SGF (described in Section \ref{sect-barrier}), MS is a model using the stationary SGF (with a large convex mesh), and MN is the Neumann model (mesh only defined in water).
The main reasons for including MS in the comparison is that this is the most well known model, and the properties we set out in the introduction refer to the ``stationary alternative''.
The main reason for including MN in the comparison is that it is the only current solution satisfying Properties 1 through 4, indeed, the only reason we do not consider this model to be a solution to the coastline problem, which would make the Barrier model superfluous, is that it introduces a new unrealistic assumption, as detailed in this section.
The complete prior model specifications, together with rationale, can be found in the supplementary material.
We also note that MS and MN are the two most commonly used models for coastlines in \texttt{R-INLA} today.
We have chosen not to compare with any of the other approaches in the literature as they do not fulfil the required properties.

\begin{figure}
	\centering
	\begin{subfigure}[b]{0.7\textwidth}
		\includegraphics[width=\linewidth]
		{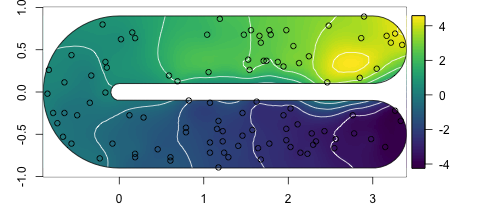}
		\caption{MS reconstruction}
		\label{fig:horseshoe-m1}
	\end{subfigure}
	
	\begin{subfigure}[b]{0.7\textwidth}
		\includegraphics[width=\linewidth]
		{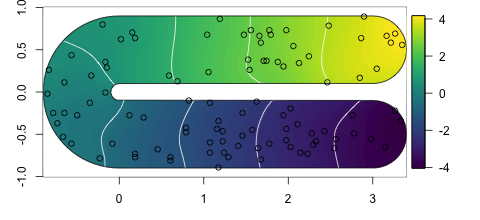}
		\caption{MB reconstruction}
		\label{fig:horseshoe-m2}
	\end{subfigure}
	
	\begin{subfigure}[b]{0.7\textwidth}
		\includegraphics[width=\linewidth]
		{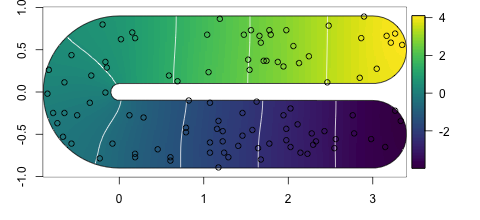}
		\caption{MN reconstruction}
		\label{fig:horseshoe-m3}
	\end{subfigure}
	
	\caption{The reconstruction estimates from the three different models.
		MS (The stationary SGF) smooths over the gap, the Barrier SGF gives a good reconstruction, and the Neumann SGF has height-curves that are orthogonal to the boundary.
	}
	\label{fig-horseshoe-models}
\end{figure}

Figure \ref{fig-horseshoe-models} shows sample reconstructions of Figure 7 for all three models, computed as the posterior mean of the SGF.
MS smooths over the gap, hence, 
in the areas near the gap the stationary reconstruction is only good
if there is an observation in a location near the boundary.
In general, if you have many observation locations along the entire boundary, the stationary reconstruction will look reasonable.
MB gives a reconstruction that seems the most similar to the true surface, including height curves that behave similarly to the true height curves.
In contrast to the true surface, MN has very straight height curves that are orthogonal to the boundary.  
Judging by these plots, the assumptions of MB seem more realistic than those of MS or MN when reconstructing an unknown function.

\begin{figure}
\centering
\includegraphics[trim=7mm 7mm 0mm 0mm,clip,width=0.9\linewidth]
{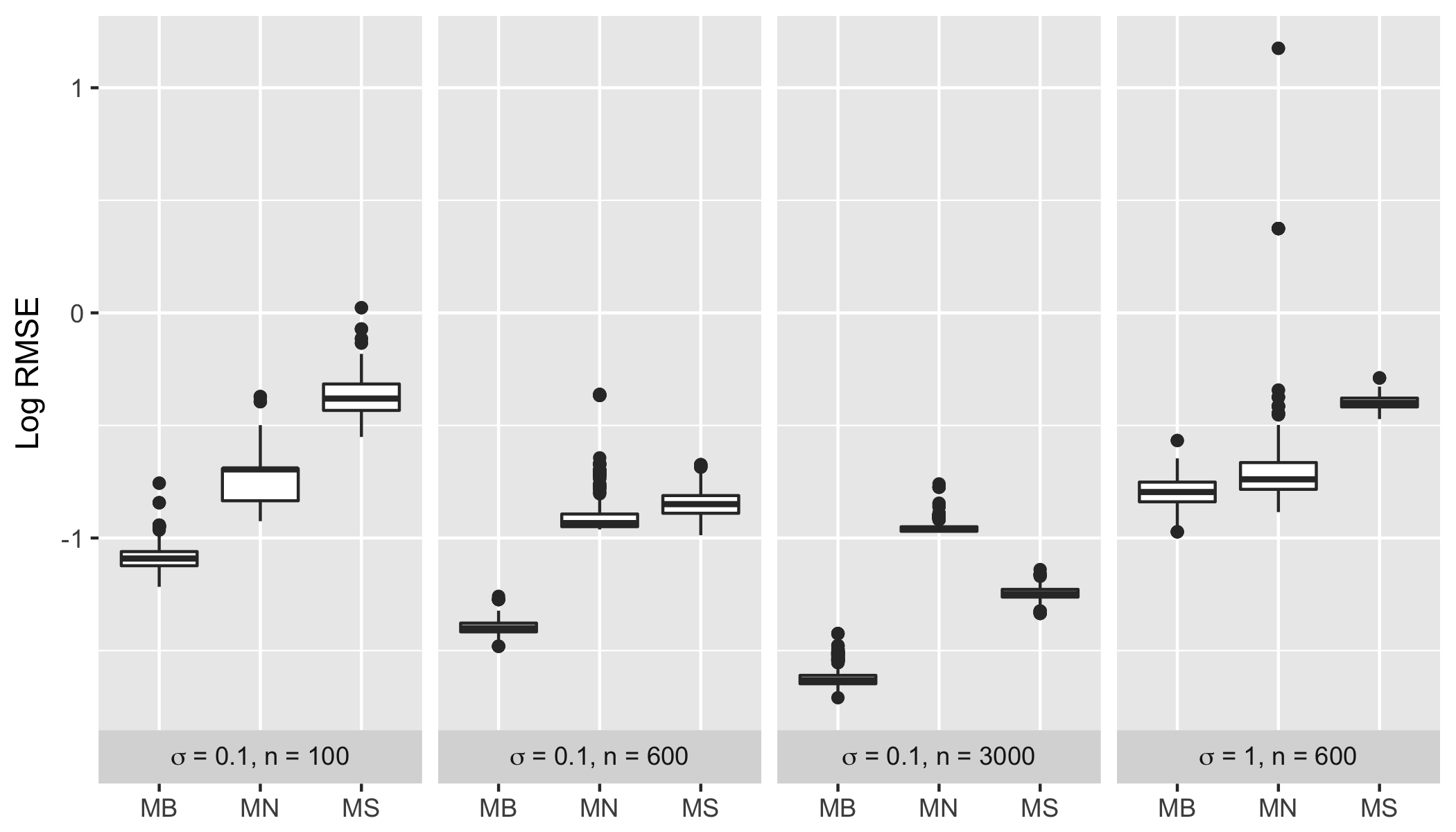}
\caption{A comparison of the RMSE when reconstructing the function in Figure \ref{fig-horseshoe-trueonly} from simulated random locations and random noise.
Figure \ref{fig-horseshoe-models} illustrates one reconstruction.
MB uses the Barrier SGF, MN the Neumann SGF, MS the stationary SGF, and the logarithm is base 10.
}
\label{fig-horseshoe-rmse-box}
\end{figure}

In the following simulation scenarios, locations were sampled uniformly at random for each run, and Gaussian noise with standard deviation $\sigma_\epsilon$ was added to the true values.
Then the three models were fitted to the data and the posterior mean of the SGF (plus intercept) used as the reconstruction.
For model comparison the root mean square error (RMSE) was used to measure the quality of the reconstruction. 
To study the variability of the reconstruction quality we simulate 1000 runs for each scenario, and show the results in Figure \ref{fig-horseshoe-rmse-box}. 
From these results we conclude that MB is the best model for reconstructing the true surface, and that this is statistically significant. 
The scale of the RMSE shows that the differences are also of large practical significance; the error of the Barrier SGF is less than half that of the stationary SGF for $\sigma_\epsilon=0.1$.
The least difference in RMSE we find is when we have a medium number of locations, and a large noise ($\sigma=1, n=600$), between the Barrier and Neumann SGFs, which 
might be due to the boundary assumption having less impact in the presence of a large measurement noise.

The simulations for $\sigma=0.1, n=3000$ are particularly interesting.
One might expect the boundary conditions to have less of an impact as $n$ increases, but that is not the case, the gain made by the Barrier SGF does not seem to diminish as $n$ increases.
Interestingly, MN is better than MS for sparse data, but MN is significantly worse for richer data.
With small sample size the informative Neumann boundary is beneficial, because this prior information contains more probability mass near the truth than the stationary prior.
In MS the prior information on boundaries is very vague and hence data, through the likelihood, can update the SGF on the boundary a lot. 
However, as sample size increases the information from data starts to overrule the prior information in MS and the posterior mean approaches the true value. 
The Neumann boundary corresponds to prior information with zero uncertainty, hence this prior is not updated by any amount of data, and performs badly for rich datasets.
In total, we conclude that the Neumann SGF does not satisfy Property 5 at all, as the results can be worse than when using a stationary SGF.


\section{Modelling fish larvae in the Finnish Archipelago}
\label{sect-archipelago}

In this application, which is the motivating example in the introduction, we model the spatial distribution fish larvae counts from the commercial fish species smelt, perch and pikeperch in the Archipelago Sea on the South-West coast of Finland.
We have 198 observations of each species.
Figure \ref{fig-archip-data} shows the study area and the observations for the smelt species, 
while a detailed description of the dataset is in appendix \ref{app-data}.
The fish larvae are sensitive to habitat changes,
therefore it is important to map the main density areas to protect the habitats when making management decisions.
The motivation for analysing this dataset from the point of view of statistical model development is that the coastline is very complex, and the data very sparse, which makes it a challenging inference problem.

\begin{figure}
	\centering
	\includegraphics[width=.7\linewidth]
	{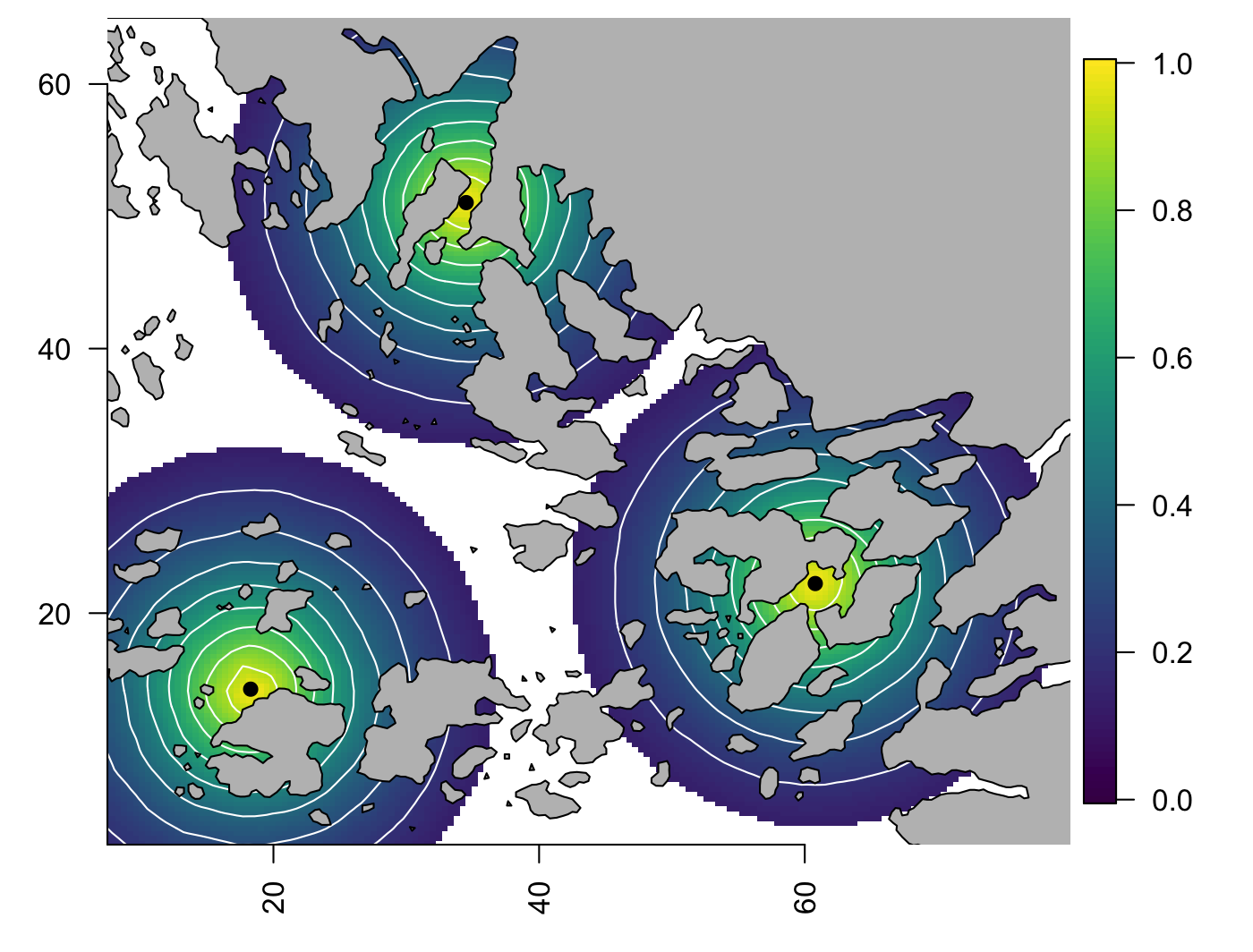}
	\includegraphics[width=.7\linewidth]
	{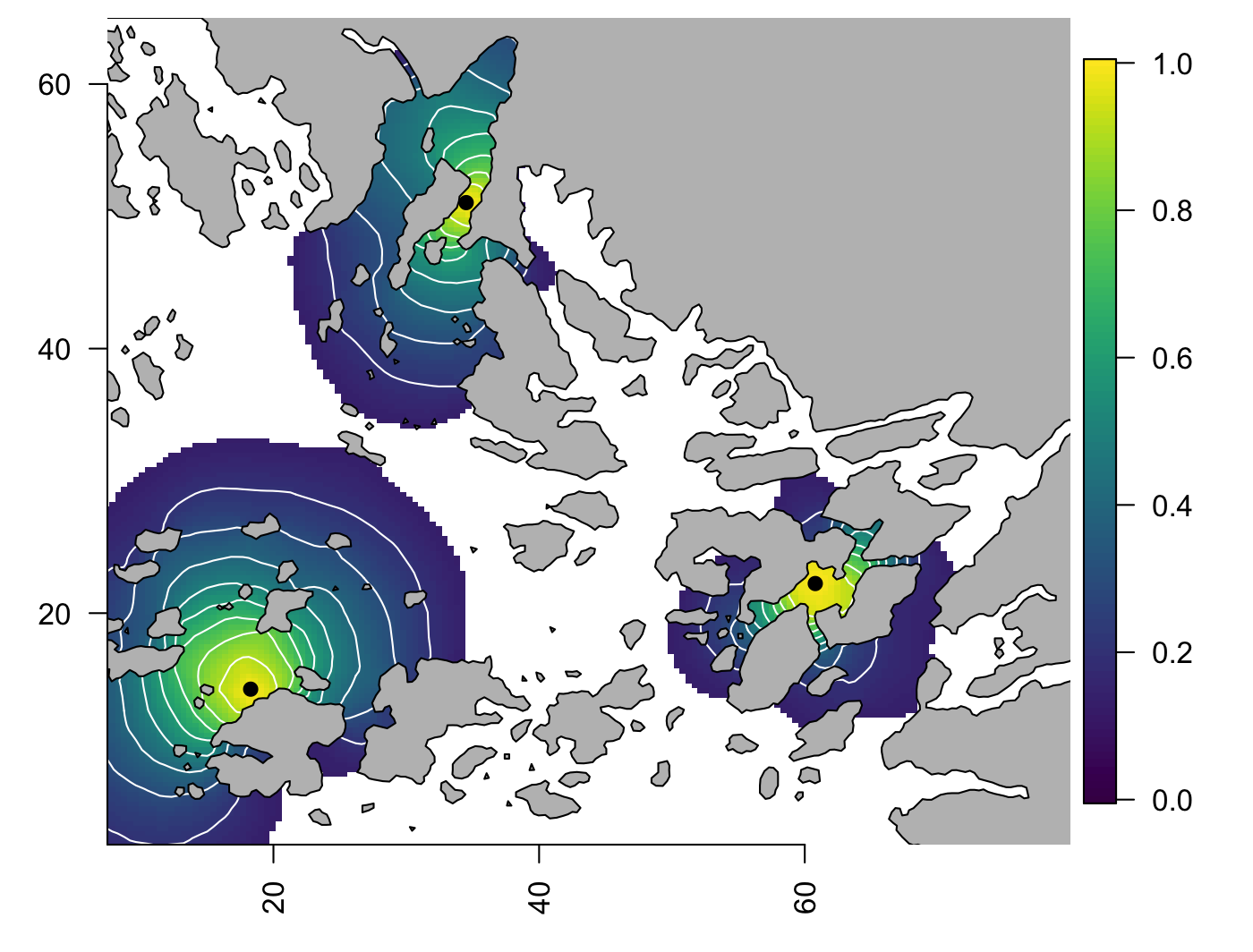}
	\caption{This figure shows the prior behaviour of the new non-stationary Barrier SGF versus the stationary alternative. 
		The map is of the Finnish Archipelago Sea, where water (white area) is considered as normal terrain, and land (grey area) is considered a physical barrier.
		The first plot shows three correlation surfaces for the stationary SGF with range $r=18\up{km}$. 
		A correlation surface is the prior correlation between any point in the plane and a chosen central point (black dot).
		The correlation was cut off at 0.1.
		The second plot shows three correlation surfaces for the Barrier SGF, using the same range $r=18\up{km}$ in the water area, and $r_b = 3.6\up{km}$ on land.
		Equidistant height curves are marked by white lines.
	}
	\label{fig-archip-prior}
\end{figure}

\begin{figure}
	\centering
	\includegraphics[trim=0.4in 0.4in 0in 0.75in,clip,width=.65\linewidth]
	{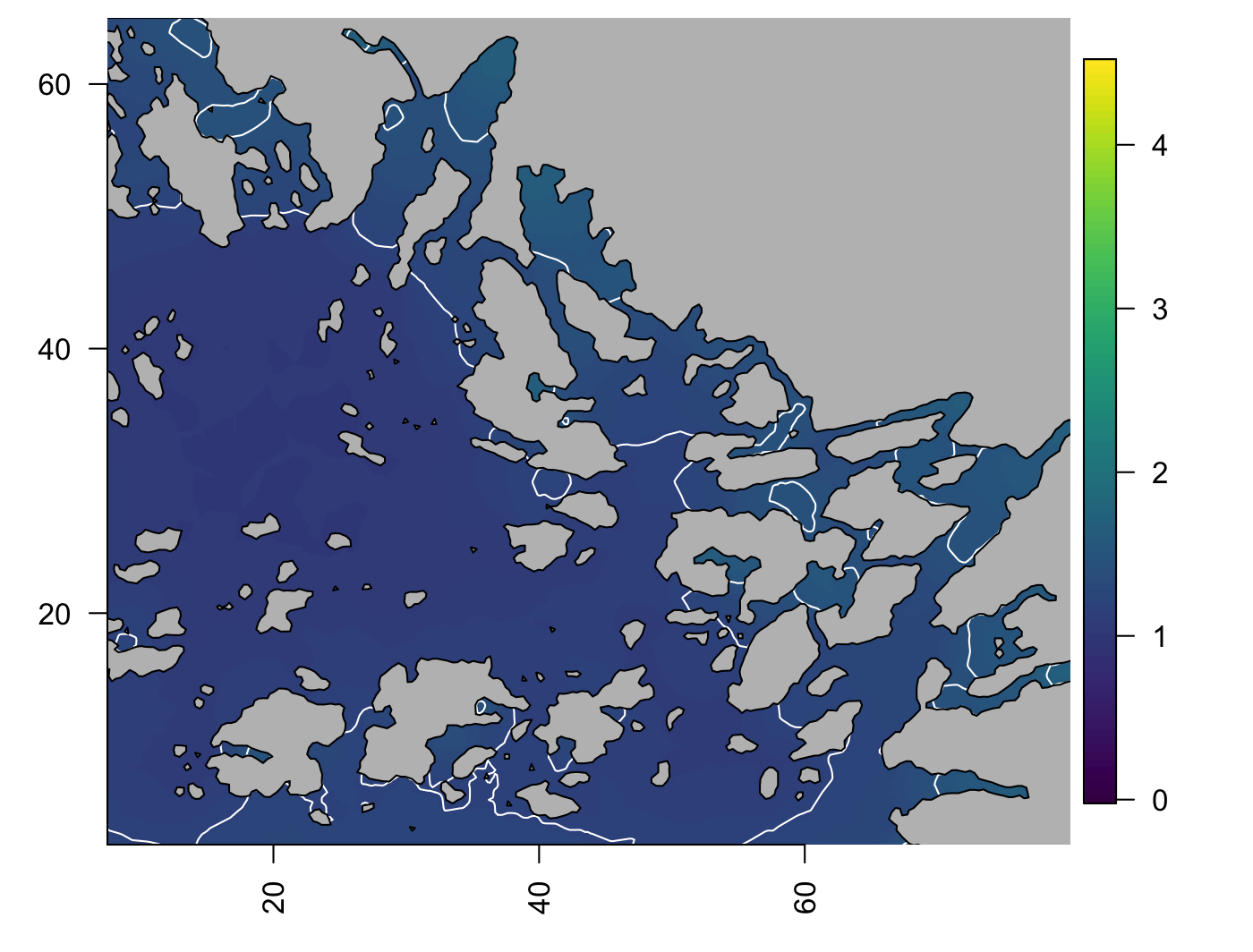}
	\caption{Example prior marginal standard deviation of the Barrier SGF, using range 18km in the water and range 3.6km on land. 
		This plot is on the same scale as we later use in Figure \ref{fig-arch-post-sd-nocov}.
	}
	\label{fig-prior-sd}
\end{figure}

We use a hierarchical Bayesian model with over-dispersed Poisson likelihood, and priors that penalise complexity (\citet{Simpson2014penalising}, \citet{Fuglstad2015interpretable}), see the supplementary material for the complete prior model specifications and rationale.
The part of the prior model we focus on here is the priors for the stationary and Barrier SGFs, see Figure \ref{fig-archip-prior} for spatial plots of the prior correlation.
From the prior correlation we see that the stationary SGF smooths over land, while the Barrier SGF smooths around land, and is hindered by the presence of large and small islands.
This is exactly the behaviour we desire from the Barrier SGF.
Figure \ref{fig-prior-sd} shows that the prior marginal standard deviation is slightly larger in narrow inlets, as discussed in Section \ref{sect-barr-finite}.
In total, we fit 6 different models to each dataset; with and without covariates, and with three different spatial components: Stationary SGF (model MS), Barrier SGF (model MB), and no spatial field (model MI).

\begin{figure}
\centering
\begin{subfigure}[b]{0.8\textwidth}
\includegraphics[width=0.9\linewidth]
{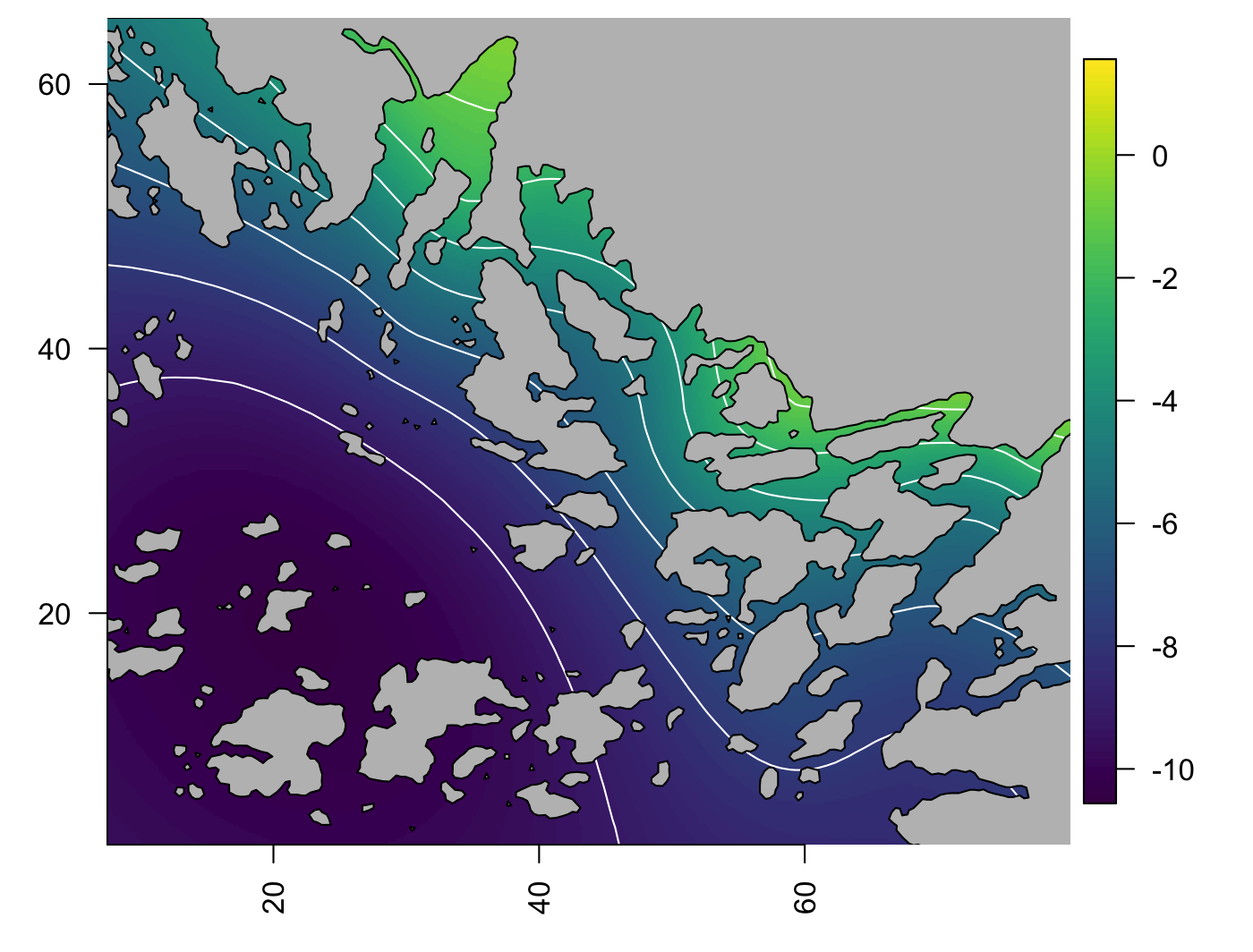}
\caption{Stationary model (MS)}
\label{fig:field01}
\end{subfigure}

\begin{subfigure}[b]{0.8\textwidth}
\includegraphics[width=0.9\linewidth]
{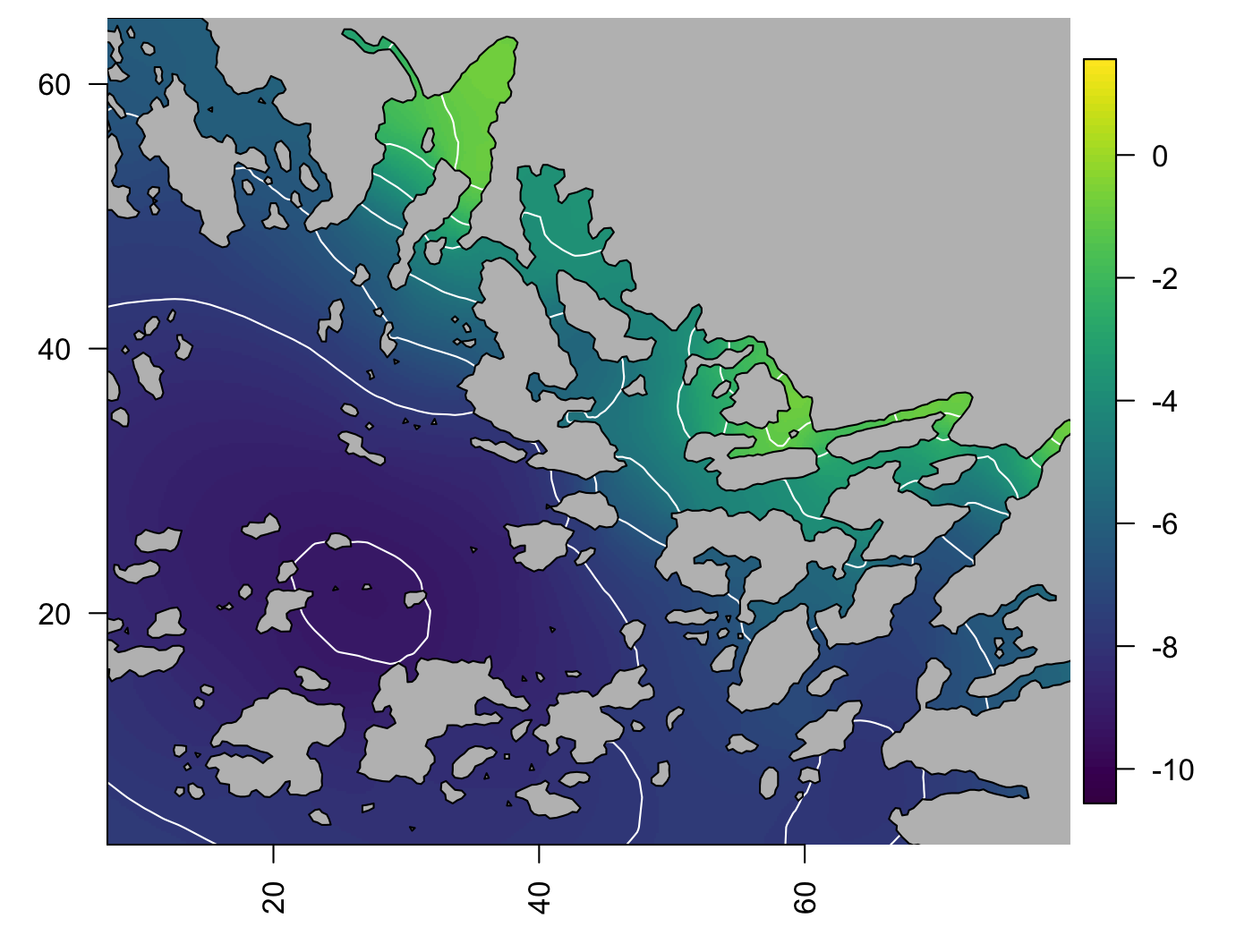}
\caption{Barrier model (MB)}
\label{fig:field02}
\end{subfigure}

\caption{Posterior mean estimate of the spatial field for the smelt larvae in the models without covariates.
We see how MS smooths over the peninsula in the northern part of the figure, while MB does not. 
}
\label{figResultSpaceNocov-1}
\end{figure}

Figure \ref{figResultSpaceNocov-1} shows the posterior mean of the spatial component, in the models for smelt larvae without covariates.
This illustrates the behaviour for datasets where there are no covariates, and datasets where the covariates are only able to model a small part of the data structure.
In the figure we see the difference between the stationary SGF and the Barrier SGF, in how they smooth across land, e.g.\ around the peninsula in the north-east.
The following Figure \ref{fig-arch-post-sd-nocov} shows the difference in the spatial uncertainty estimation.
In the south-east part of these plots we find inlets where we get higher uncertainty in the Barrier SGF compared to the stationary SGF. 
This is in accordance with our intuition, as we do not have much information about what happens there, because the coastline is separating the inlets from nearby measurement locations.

Figure \ref{fig:ResultCovars} summarizes the point estimates and 95\% credible intervals for the fixed effects in the models of smelt larvae.
The posterior mean and intervals change from a model with no SGF to any of the spatial models, because
the covariates are all spatially structured, hence confounded with the spatial model component.
This confounding is desirable as it accounts for unmeasured covariates and spatial noise, making the fixed effects more reliable for extrapolation to other regions or other studies.
We also observe changes in the posterior intervals between the model with stationary SGF (MS) and the model with Barrier SGF (MB). 
These changes may not be large enough to motivate the use of a very resource intensive non-stationary SGF, but 
considering the ease of use of the Barrier SGF, they are large enough that we recommend this non-stationary SGF for future applications.
For completion, we show the results for the perch and the pikeperch larvae datasets in the supplementary material.

To investigate whether any of the datasets have a clear preference for one of the models, we compare all 6 models for each of the 3 species.
The comparison criterion we use is Leave-One-Out Cross-Validation (LOOCV) with negative log predictive density (NLPD), and we bootstrap the mean score differences to understand the variability.
A simpler scoring method would give us an estimate of which model is best, but would be less reliable.
We compute the score by running each model 198 times for each dataset, see the supplementary material.
The results of this model comparison are almost all inconclusive, except for the result that 
the models with only intercept and iid effect are bad, all other models are approximately equally good, as the zero level is within the variability of the comparison criterion.


We now argue that property 3 is satisfied by the Barrier SGF.
From the LOOCV, together with the inference performed in Section \ref{sect-horseshoe} and extensive informal experimentation, we conclude that models based on the Barrier SGF are numerically stable, 
as we have not encountered cases where a model with the Barrier SGF failed to converge but one with the stationary SGF succeeded.
A simulation study in the archipelago can be found in the supplementary material.


\begin{figure}
	\centering 
	\begin{subfigure}[b]{0.8\textwidth}
		\includegraphics[width=0.9\linewidth]
		{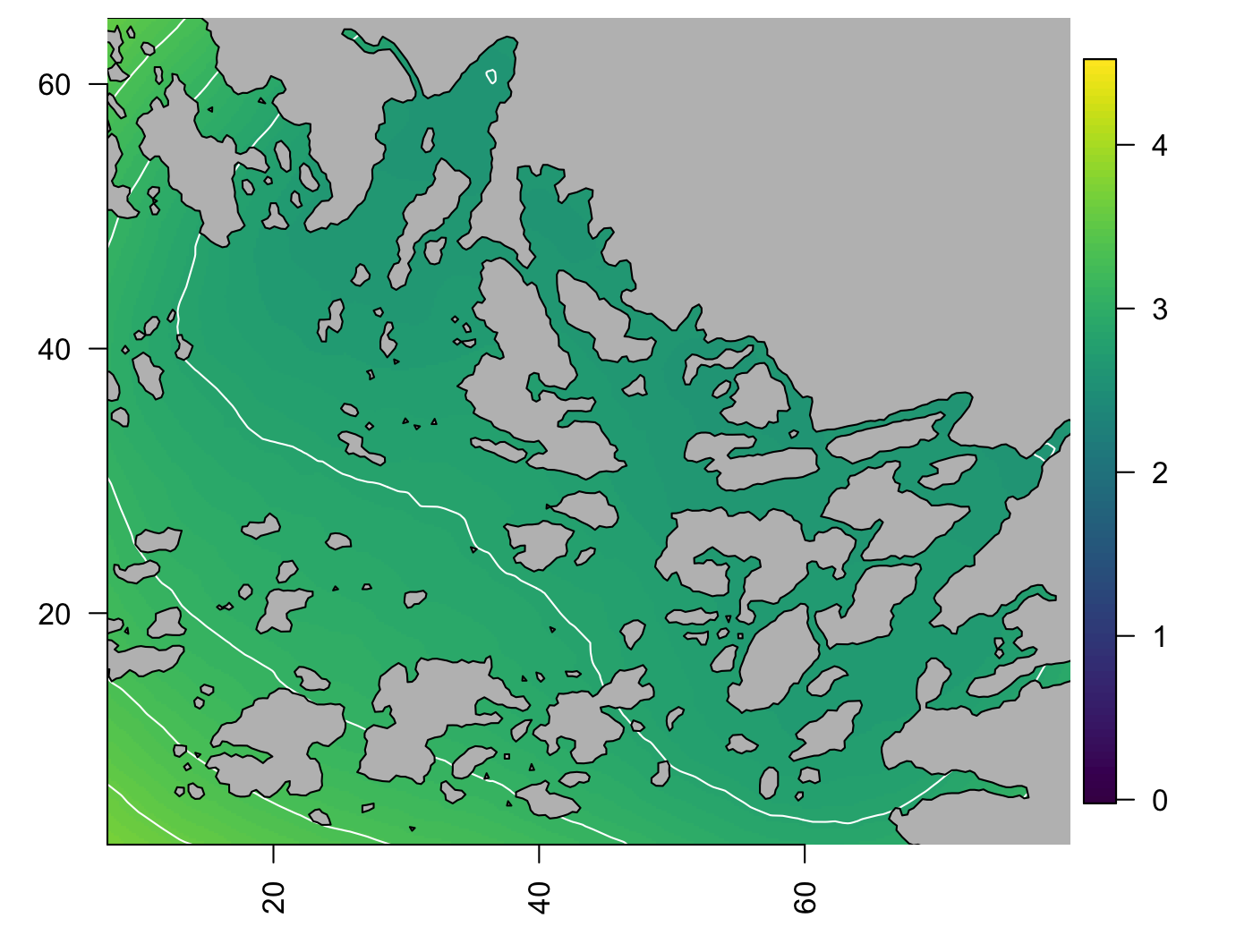}
		\caption{Stationary model (MS)}
		\label{fig:field03}
	\end{subfigure}
	
	\begin{subfigure}[b]{0.8\textwidth}
		\includegraphics[width=0.9\linewidth]
		{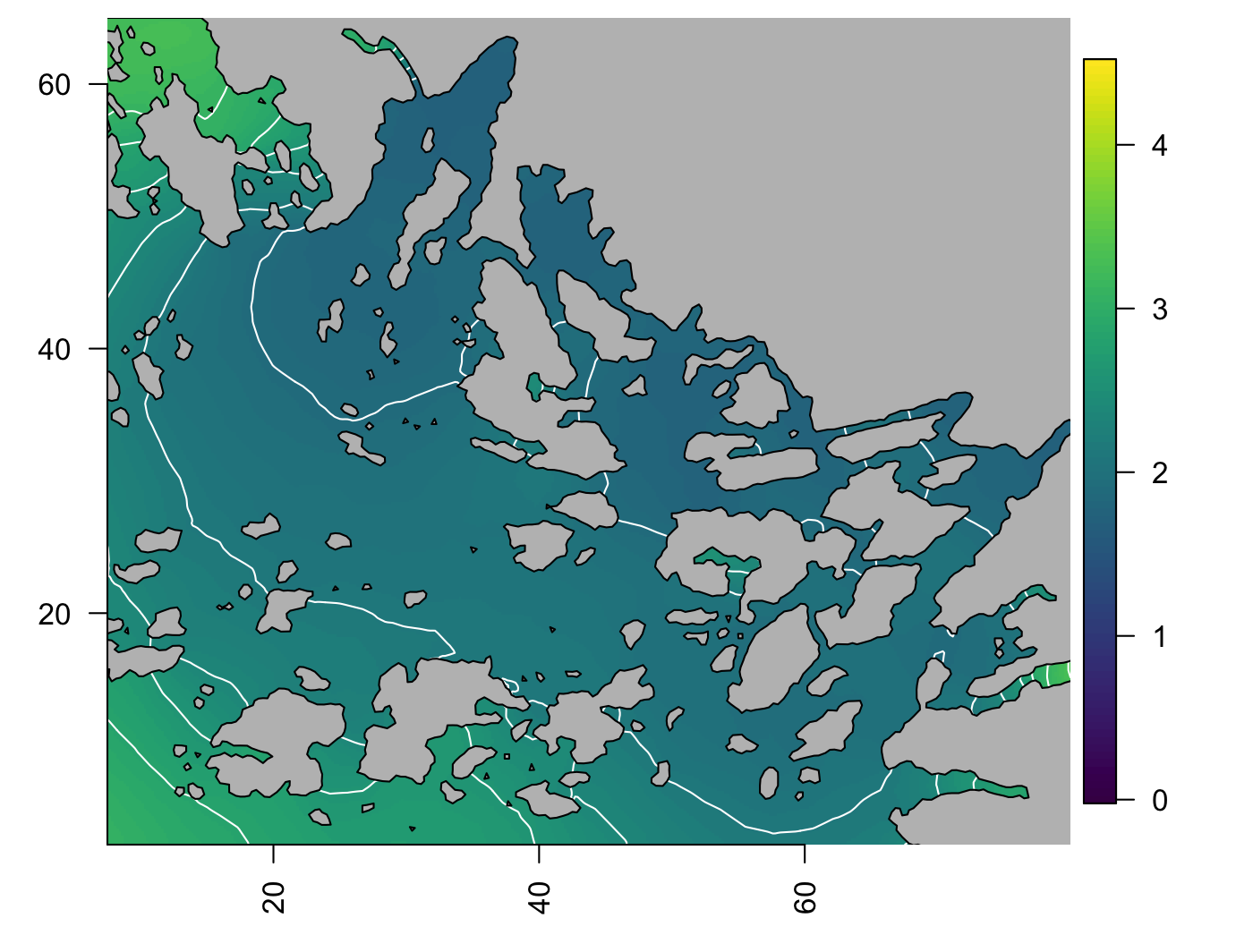}
		\caption{Barrier model (MB)}
		\label{fig:field04}
	\end{subfigure}
	
	\caption{Posterior spatial uncertainty (marginal standard deviation) for the smelt larvae in the models without covariates.
		We see that the uncertainty in MB is lower over all, but larger in inlets.
	}
	\label{fig-arch-post-sd-nocov}
\end{figure}

\begin{figure}
\centering
\includegraphics[trim=7mm 7mm 0mm 0mm,clip,width=0.9\linewidth]
{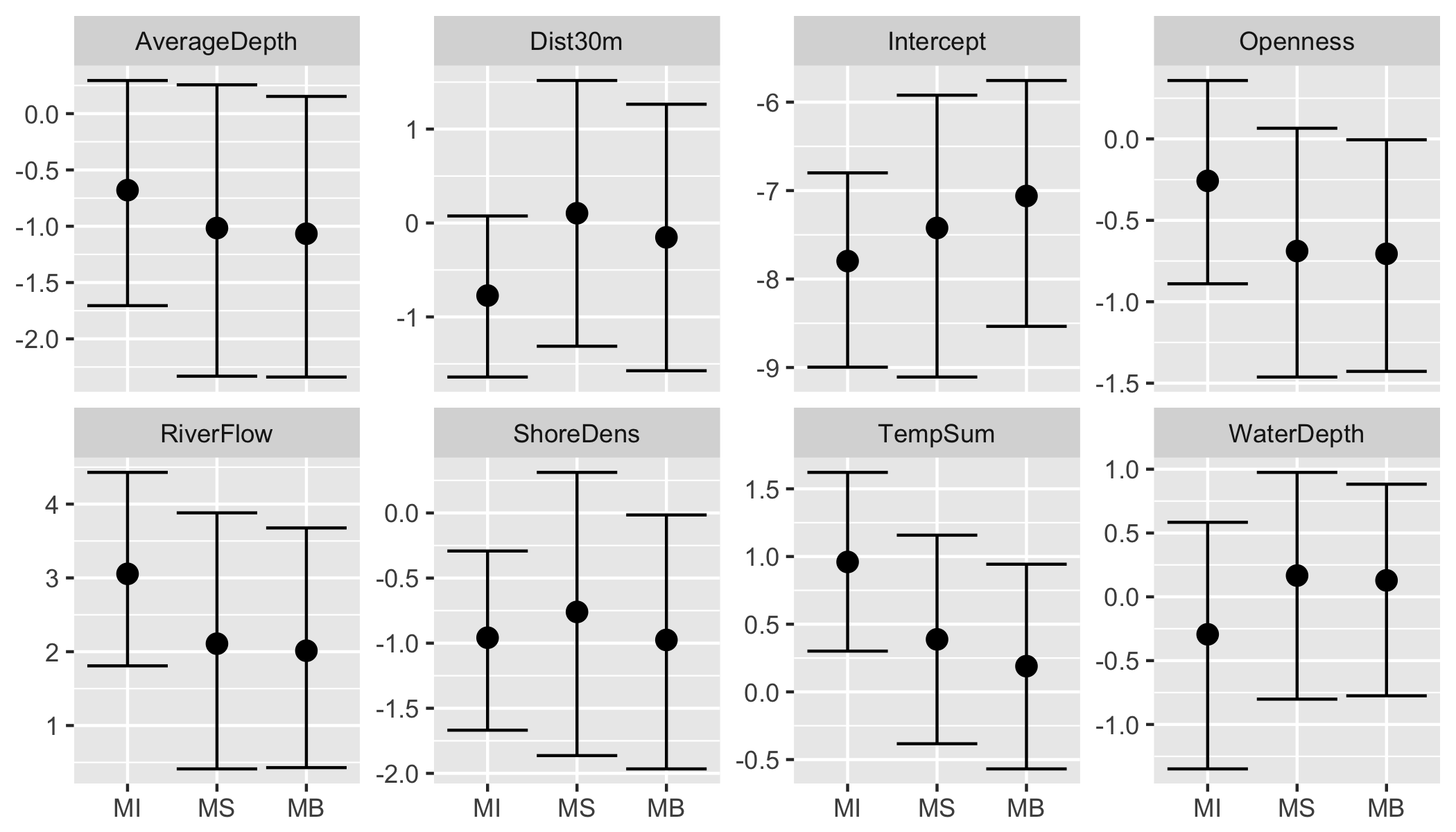}
\caption{A comparison of the posterior medians and 95\% credible intervals for the fixed effects in the smelt analysis between three models. 
MI is the model without spatial field, MS the model with stationary SGF, and MB the model with Barrier SGF.
}
\label{fig:ResultCovars}
\end{figure}

\section{Discussion}

In statistics, one of the main aims is to construct realistic models that are motivated by common sense and scientific knowledge. In spatial modelling, classical models are unrealistic when they smooth over holes or physical barriers, and thus, to address these issues, statisticians have developed several more realistic models.  However, realism alone may not be sufficient; if the improved models have other weaknesses, compared to the classical models, they may not always be useful. So far, there seems to be a trade-off, in that more realistic models are much more time consuming to fit and/or more complex to use, and/or introduce new unrealistic assumptions, compared to the classical models. The achievement in this paper is that we developed a new realistic approach, the Barrier model, that does not have this trade-off. As highlighted by the five properties we defined, the new model seems to have no practical weaknesses. Therefore, users can use it in all application scenarios, including those with sparse data and very complex boundaries.

%


The primary advantage of the Barrier SGF compared to previous approaches is the computational efficiency.
Through the SPDE formulation, we achieve a computational cost equal to the stationary model, in theory, because we obtain the same sparsity structure for the precision matrix of the Barrier SGF as for the stationary SGF.
The current implementation of the Barrier model is in R, and the stationary model is in C, leading to different computational times between the two models when using the code provided in the supplementary material.
This efficiency allows us to perform thousands of simulation-inference runs in Section \ref{sect-horseshoe}, 
a cross-validation study for the Archipelago application (see supplementary material), and a simulation study for the Archipelago (see supplementary material).
Further, using these precision matrices in \texttt{R-INLA} is fast for datasets 
with a hundred thousand observations, 
and allows us to construct space-time models defined as Kronecker products of the 
Barrier SGF and a temporal model.

In this paper we also discuss whether the inference results are different enough, between the stationary model and the Barrier model, to recommend using the model in practice.
For reconstructing the horseshoe test function, we get large differences, similarly to the existing literature (e.g.\ \citet{Wood2008Soapfilm}).
For this example it is clear that neither the stationary nor the Neumann model can be used, and an improved model is needed.
However, we caution against comparing advanced approaches to the coastline problem by their performance on this single test function.
Similarly to how the original test function preferred models with Neumann boundary assumption \citep{Wood2008Soapfilm}, the current Horseshoe prefers a specific type of models.
The ``winner'' of a comparison based on any single example would mostly be determined by which approach has the most similar prior (or null-space, or ``low penalty space'') to the true boundary of the Horseshoe function.
Comparing different approaches to the coastline problem must be done with greater care, and we would like to see defined a collection of different test functions, with different coastlines, satisfying very different boundary conditions, to get a more general comparison criterion.

In the Archipelago application, the field is sparsely observed, and so, the posterior SGF changes substantially between the stationary SGF and the Barrier SGF.
When visualising (smoothing) spatial data, scientists will not accept a visualisation that is intuitively inappropriate, as is the case with the result of the stationary SGF.
In all applications where the spatial estimation maps are important we recommend using the Barrier model.

For applications where the main concern is estimating the coefficients of the fixed effect (or similar non-linear terms), and the SGF is a nuisance parameter, the difference between using a simple model and a model taking the coastline into account may be almost negligible.
Based on the results in this paper, and experimentation with two other datasets, we cannot in general 
recommend the use of a complex and time consuming approach to the coastline problem.
However, we can recommend using the Barrier model, as the time and effort required to fit this model is almost the same as for the stationary model.

The weaker points of the results in this paper are the arbitrariness of the choice of range fraction $r_b/r$, and the inability to deal with physical barriers that are infinitely thin.
Ad-hoc sensitivity analysis showed that the fraction we chose had almost no impact on parameter estimation.
The simulation study in the Archipelago (see supplementary material) showed that the choice of range fraction did not impact predictive performance.
Furthermore, any concerned user can study the impact of this choice through plotting prior correlation surfaces, \emph{before} any data is fitted.
As a side note, since the impact of changing this parameter (the range fraction, alternatively $r_b$) 
is so small, we strongly advise against attempting to estimate it.
For dealing with physical barriers that are infinitely thin, we suggest making the barriers artificially thicker, so that the width is at least one mesh triangle across, and to use a very small range fraction.

There are several avenues for future research.
First, the equations we solved in Section 3 can be used to infer $k$ parameters in a non-stationary model, and we are looking into whether there are interesting applications of such an approach.
Second, the SPDE approach can be used to model time-varying physical barriers in a space-time model, for example when regions are closed for fishing to preserve species.
Third, the discussion of boundary conditions can be carried out into the field of SAR/CAR models in general, for example in diseasemapping.

We end with a small remark; the Barrier SGF is in one view more flexible than the stationary SGF, as its correlation function (from empirical evidence) is always approximately the same or lower than that of the stationary SGF; the Barrier SGF may decouple some observations compared to the stationary SGF, but it does not increase other dependencies, when using the same range $r$ in both models.

\section{Supplementary materials}
The Barrier SGF is implemented in the \object{R} package \package{INLA}, as \texttt{inla.barrier.pcmatern},
and the main step needed, compared to fitting a stationary model, 
is to specify which part of the study area is the physical barrier, e.g.\ by a polygon. 
See the code examples at \url{https://haakonbakka.bitbucket.io/btopic107.html} and \\ \url{https://haakonbakka.bitbucket.io/btopic110.html}. 
For an example on how to construct an appropriate mesh, see \url{https://haakonbakka.bitbucket.io/btopic104.html}.
[After review: Attach a zip with the website and one with the full code used in the paper.]

Refer to the supplementary material for additional details.

\section{Acknowledgements}
We are grateful to Simon Wood
and Rosa Crujeiras Casais
for detailed feedback on this research project,
to Finn Lindgren for assistance with understanding the finer details of the SPDE approach,
and to David Bolin for assistance with the theory of existence of solutions for SPDEs.
Data collection was funded by VELMU and the Natural Resources Institute Finland (Luke).
We appreciate the detailed feedback from reviewers.

\bibliography{biblio-hb-18-01-09-plus}

\begin{thebibliography}{}

\bibitem[Bakka, 2018]{bakka2018solve}
Bakka, H. (2018).
\newblock How to solve the stochastic partial differential equation that gives
  a {M}at\'ern random field using the finite element method.
\newblock {\em arXiv preprint arXiv:1803.03765}.

\bibitem[Bergstr{\"o}m et~al., 2014]{Sundblad+etal:2014}
Bergstr{\"o}m, U., Sundblad, G., Sandstr{\"o}m, A., and Ekl{\"o}v, P. (2014).
\newblock Nursery habitat availability limits adult stock sizes of predatory
  coastal fish.
\newblock {\em ICES Journal of Marine Science}, 71:672--680.

\bibitem[Bhatt et~al., 2015]{art618}
Bhatt, S., Weiss, D.~J., Cameron, E., Bisanzio, D., Mappin, B., Dalrymple, U.,
  Battle, K.~E., Moyes, C.~L., Henry, A., Eckhoff, P.~A., Wenger, E.~A.,
  Briët, O., Penny, M.~A., Smith, T.~A., Bennett, A., Yukich, J., Eisele,
  T.~P., Griffin, J.~T., Fergus, C.~A., Lynch, M., Lindgren, F., Cohen, J.~M.,
  Murray, C. L.~J., Smith, D.~L., Hay, S.~I., Cibulskis, R.~E., and Gething,
  P.~W. (2015).
\newblock The effect of malaria control on plasmodium falciparum in {A}frica
  between 2000 and 2015.
\newblock {\em Nature}, (526):207--211.

\bibitem[Blangiardo and Cameletti, 2015]{blangiardo2015spatial}
Blangiardo, M. and Cameletti, M. (2015).
\newblock {\em Spatial and spatio-temporal Bayesian models with R-INLA}.
\newblock John Wiley \& Sons.

\bibitem[Bolin and Kirchner, 2018]{Bolin17rational}
Bolin, D. and Kirchner, K. (2018).
\newblock The rational {SPDE} approach for {G}aussian random fields with
  general smoothness.
\newblock {\em arXiv preprint arXiv:1711.04333}.

\bibitem[Bolin et~al., 2017]{bolin2017numerical}
Bolin, D., Kirchner, K., and Kov{\'a}cs, M. (2017).
\newblock Numerical solution of fractional elliptic stochastic pdes with
  spatial white noise.
\newblock {\em arXiv preprint arXiv:1705.06565}.

\bibitem[Diggle, 2010]{diggle2010historical}
Diggle, P.~J. (2010).
\newblock Historical introduction.
\newblock In Gelfand, A., Diggle, P., Fuentes, M., and Guttorp, P., editors,
  {\em Handbook of Spatial Statistics}, pages 3--16. CRC/Chapman \& Hall, Boca
  Raton, FL.

\bibitem[Fuglstad et~al., 2017]{Fuglstad2015interpretable}
Fuglstad, G.-A., Simpson, D., Lindgren, F., and Rue, H. (2017).
\newblock Constructing priors that penalize the complexity of gaussian random
  fields.
\newblock {\em Journal of the American Statistical Association},
  (just-accepted).

\bibitem[Gander and Dubois, 2015]{gander2015optimized}
Gander, M.~J. and Dubois, O. (2015).
\newblock Optimized schwarz methods for a diffusion problem with discontinuous
  coefficient.
\newblock {\em Numerical Algorithms}, 69(1):109--144.

\bibitem[Golding et~al., 2017]{golding2017mapping}
Golding, N., Burstein, R., Longbottom, J., Browne, A.~J., Fullman, N.,
  Osgood-Zimmerman, A., Earl, L., Bhatt, S., Cameron, E., Casey, D.~C., et~al.
  (2017).
\newblock Mapping under-5 and neonatal mortality in {A}frica, 2000--15: a
  baseline analysis for the sustainable development goals.
\newblock {\em The Lancet}, 390(10108):2171--2182.

\bibitem[Grisvard, 1985]{grisvard1985elliptic}
Grisvard, P. (1985).
\newblock Elliptic problems in nonsmooth domains, volume 24 of monographs and
  studies in mathematics. {Pitman}.

\bibitem[Illian et~al., 2012]{illian2012toolbox}
Illian, J.~B., S{\o}rbye, S.~H., and Rue, H. (2012).
\newblock A toolbox for fitting complex spatial point process models using
  integrated nested {Laplace} approximation ({INLA}).
\newblock {\em The Annals of Applied Statistics}, pages 1499--1530.

\bibitem[Kallasvuo et~al., 2017]{Kallasvuo+etal:submitted}
Kallasvuo, M., Vanhatalo, J., and Veneranta, L. (2017).
\newblock Modeling the spatial distribution of larval fish abundance provides
  essential information for management.
\newblock {\em Canadian Journal of Fisheries and Aquatic Sciences},
  74:636--649.

\bibitem[Lindgren et~al., 2011]{Lindgren2011}
Lindgren, F., Rue, H., and Lindstr\"{o}m, J. (2011).
\newblock An explicit link between {Gaussian} fields and {Gaussian} {Markov}
  random fields: the stochastic partial differential equation approach.
\newblock {\em Journal of the Royal Statistical Society: Series B (Statistical
  Methodology)}, 73(4):423--498.

\bibitem[Miller and Wood, 2014]{miller2014finite}
Miller, D.~L. and Wood, S.~N. (2014).
\newblock Finite area smoothing with generalized distance splines.
\newblock {\em Environmental and ecological statistics}, 21(4):715--731.

\bibitem[Noor et~al., 2014]{art615}
Noor, A.~M., Kinyoki, D.~K., Mundia, C.~W., Kabaria, C.~W., Mutua, J.~W.,
  Alegana, V.~A., Fall, I.~S., and Snow, R.~W. (2014).
\newblock The changing risk of {P}lasmodium falciparum malaria infection in
  {A}frica: 2000-10: a spatial and temporal analysis of transmission intensity.
\newblock {\em The Lancet}, 383(9930):1739--1747.

\bibitem[Ramsay, 2002]{Ramsay2002Spline}
Ramsay, T. (2002).
\newblock Spline smoothing over difficult regions.
\newblock {\em Journal of the Royal Statistical Society: Series B (Statistical
  Methodology)}, 64(2):307--319.

\bibitem[Rue et~al., 2009]{Rue2009}
Rue, H., Martino, S., and Chopin, N. (2009).
\newblock Approximate bayesian inference for latent gaussian models by using
  integrated nested laplace approximations.
\newblock {\em Journal of the Royal Statistical Society: Series B (Statistical
  Methodology)}, 71(2):319--392.

\bibitem[Rue et~al., 2017]{Rue2017bayesian}
Rue, H., Riebler, A., S{\o}rbye, S.~H., Illian, J.~B., Simpson, D.~P., and
  Lindgren, F.~K. (2017).
\newblock Bayesian computing with {INLA}: a review.
\newblock {\em Annual Review of Statistics and Its Application}, 4:395--421.

\bibitem[Sangalli et~al., 2013]{sangalli2013spatial}
Sangalli, L.~M., Ramsay, J.~O., and Ramsay, T.~O. (2013).
\newblock Spatial spline regression models.
\newblock {\em Journal of the Royal Statistical Society: Series B (Statistical
  Methodology)}, 75(4):681--703.

\bibitem[Scott-Hayward et~al., 2014]{scott2014complex}
Scott-Hayward, L. A.~S., MacKenzie, M.~L., Donovan, C.~R., Walker, C., and
  Ashe, E. (2014).
\newblock Complex region spatial smoother ({CReSS}).
\newblock {\em Journal of Computational and Graphical Statistics},
  23(2):340--360.

\bibitem[Shpilev et~al., 2005]{Shpilev+etal:2005}
Shpilev, H., Ojaveer, E., and Lankov, A. (2005).
\newblock Smelt (\emph{Osmerus eperlanus L.}) in the {Baltic} {Sea}.
\newblock {\em Proceedings of the Estonian Academy of Sciences, Biology and
  Ecology}, 54:230--241.

\bibitem[Simpson et~al., 2017]{Simpson2014penalising}
Simpson, D., Rue, H., Riebler, A., Martins, T.~G., and S{\o}rbye, S.~H. (2017).
\newblock Penalising model component complexity: A principled, practical
  approach to constructing priors.
\newblock {\em Statistical Science}, 32(1):1--28.

\bibitem[Vanhatalo et~al., 2012]{Vanhatalo+etal:2012}
Vanhatalo, J., Veneranta, L., and Hudd, R. (2012).
\newblock Species distribution modelling with gaussian processes: a case study
  with the youngest stages of sea spawning whitefish (\emph{Coregonus lavaretus
  L. s.l.}) larvae.
\newblock {\em Ecological Modelling}, 228(0):49 -- 58.

\bibitem[Wang and Ranalli, 2007]{wang2007low}
Wang, H. and Ranalli, M.~G. (2007).
\newblock Low-rank smoothing splines on complicated domains.
\newblock {\em Biometrics}, 63(1):209--217.

\bibitem[Whittle, 1954]{Whittle1954}
Whittle, P. (1954).
\newblock On stationary processes in the plane.
\newblock {\em Biometrika}, 41(3/4):pp. 434--449.

\bibitem[Wood et~al., 2008]{Wood2008Soapfilm}
Wood, S.~N., Bravington, M.~V., and Hedley, S.~L. (2008).
\newblock Soap film smoothing.
\newblock {\em Journal of the Royal Statistical Society: Series B (Statistical
  Methodology)}, 70(5):931--955.

\end{thebibliography}


\begin{thebibliography}{}

\bibitem[Fuglstad et~al., 2017]{Fuglstad2015interpretable}
Fuglstad, G.-A., Simpson, D., Lindgren, F., and Rue, H. (2017).
\newblock Constructing priors that penalize the complexity of gaussian random
  fields.
\newblock {\em Journal of the American Statistical Association},
  (just-accepted).

\bibitem[Lind{\'e}n and M{\"a}ntyniemi., 2011]{Linden+Mäntyniemi:2011}
Lind{\'e}n, A. and M{\"a}ntyniemi., S. (2011).
\newblock Using the negative binomial distribution to model overdispersion in
  ecological count data.
\newblock {\em Ecology}, 92:1414--1421.

\bibitem[Simpson et~al., 2017]{Simpson2014penalising}
Simpson, D., Rue, H., Riebler, A., Martins, T.~G., and S{\o}rbye, S.~H. (2017).
\newblock Penalising model component complexity: A principled, practical
  approach to constructing priors.
\newblock {\em Statistical Science}, 32(1):1--28.

\end{thebibliography}
\appendix

\section{Proof of theorem \ref{theorem-existance}} 
\label{app-existence-proof}

The operator $L$ is clearly linear, self-adjoint and positive definite.
From the decay of the eigenvalues we get that $L$ inverse is the limit of finite rank operators, hence $L$ has a compact inverse.
In total, $L$ satisfies all the properties required for the results in \citet{bolin2017numerical}.
Define $\dot H^{-r}$ to be the dual space of $\dot H^r$ with respect to the inner product on the space $H$.
From Lemma 2.1 in \citet{bolin2017numerical} there is a unique continuous extension of $L$ to
an isometric isomorphism from $\dot H^s$ to $\dot H^{s-2}$ for any $s$.
Proposition 2.3 and Remark 2.4 detail how this proves existence and uniqueness of a solution of the SPDE
$$ L u = \m W.$$
Further, for any $\epsilon >0 $ we get $u \in L_2(\Pi; \dot H^{1-\epsilon})$.
In particular, $u \in L_2(\Pi; H)$.

We note that the results in \citet{bolin2017numerical} can also be used to study the behaviour of the FEM approximation.

\section{The fish larvae dataset}

\label{app-data}

In this appendix we provide a more detailed description of the fish larvae dataset.

The study area is located in the Archipelago Sea on the south-west coast of Finland in the northern Baltic Sea (Figure \ref{fig-archip-data}). 
The Baltic Sea is one of the largest brackish water bodies in the world consisting of shallow, topographically complex and extensive archipelago rich in islands. 
Environmental gradients are typically strong, both north-southward and west-eastward along the coastline but also from inshore to offshore, e.g. spring-time temperature sum (see table \ref{tab-covar}) and turbidity vary strongly between inner bays and open water area due to the influence of river runoff. 
The archipelago and coastal areas host many essential biological processes such as fish reproduction.
Hence, knowledge on the specific reproduction areas is of central importance in marine spatial planning and fisheries management \citep{Vanhatalo+etal:2012,Kallasvuo+etal:submitted}.

As a case study we consider three species of fish, smelt (\emph{Osmerus eperlanus}), perch (\emph{Perca fluviatilis}) and pikeperch (\emph{Sander lucioperca}). 
They are of freshwater origin and spawn in shallow coastal waters in low salinity estuaries and river mouths in the northern Baltic Sea  \citep{Shpilev+etal:2005,Sundblad+etal:2014}. 
They are fished commercially and are also highly sought after by recreational fishers. 
Our main interest is in the early-stage larvae, which are found relatively close to the spawning sites. 
We use a subset of data (198 sampling locations) introduced by \citet{Kallasvuo+etal:submitted}. 
The data were collected in 2007 and 2011 and comprise of the number of larvae per sampling location together with information on varying sampling effort (measured as the volume of water sampled). 
The environmental variables included six variables, which were available in GIS format in 50m resolution throughout the study area, and spatial coordinates (see Table \ref{tab-covar}). 
Sampling year was also available, but not included, as there were only two unique values, and the exploratory data analysis did not indicate any differences between years.
If the study had contained more years, it would have been natural to include a temporal model component in the model, as in \citet{Kallasvuo+etal:submitted}. 

The covariates, except for \joetdsumsqNospace, were standardized to have mean zero and variance 1 before the analysis.
This was mainly done in order to facilitate interpretation of the estimates and the uncertainty of the fixed effects.  This also implies that due to standardization the priors act in the same way on all fixed effects. 
The response of smelt abundance along the covariate \joetdsumsq was step-like in the analysis of \citet{Kallasvuo+etal:submitted}, which comprised the whole coastline of Finland. 
In our sub-area, 193 sites out of all 198 sites took on a total of only two different values. 
It is unlikely that a meaningful linear relationship can be derived given these values, and hence, we recoded the covariate as a factor covariate in this study, where 0 represents the values below the average and 1 represents values greater than the average.



\begin{table}
\centering
\begin{tabular}{|c|p{0.57\textwidth}|}
\hline \textbf{Name} & \textbf{Covariate description} \\
\hline \dptavg & Average depth in a circle of 15 km; describes the water depth gradient in a large spatial scale.\\
\hline \dist & Distance to 30 m (or more) depth zone; implicates location in the archipelago so that, e.g. sheltered inner bays are emphasized by this covariate. \\ 
\hline \joetdsumsq & Square root of inverse distance to nearest river mouth weighted with annual average runoff; describes the influence of the river mouths and freshwater runoff. \\ 
\hline lined15km & Shoreline length in a circle of 15 km; describes the effect of wind exposure and water exchange. \\ 
\hline \swmlog & $\log_{10}$ of wave exposure; describes the degree of wave exposure \\ 
\hline \temjul & Cumulative temperature sum from ice-break to July 15; describes how rapidly water area warms up in spring after ice break-up. \\ 
\hline 
\end{tabular} 
\caption{Covariates used in the models. }
\label{tab-covar}
\end{table}

\section{A small code example}

\label{app-codesnip}

In this appendix we provide a small code example to show the implementation needed by users of the Barrier SGF.
The relevant part of the code for MS was \\
\indent \texttt{
	>> spde = inla.spde2.pcmatern(mesh,
} \\
\indent \indent \indent \texttt{prior.range = c(6, .5), prior.sigma = c(3, 0.01))} \\
\indent \texttt{	
	>> formula = y~ -1+m + f(s, model=spde) + 
} \\
\indent \indent \indent \texttt{f(iidx, model="iid", hyper=hyper.iid)} \\
while for MB we replace that code with \\
\indent \texttt{
	>> barrier.model = inla.barrier.pcmatern(mesh, barrier.triangles, 
} \\
\indent \indent \indent \texttt{prior.range = c(6, .5), prior.sigma = c(3, 0.01))} \\
\indent \texttt{
	>> formula = y~ -1+m + f(s, model=barrier.model) + 
} \\
\indent \indent \indent \texttt{f(iidx, model="iid", hyper=hyper.iid)} \\
where \texttt{barrier.triangles} is a list of indices of the mesh triangles covering land.


\clearpage

\end{document}




\section{Supplementary material}


\subsection{The prior model for the horseshoe reconstruction}
The simulation, and inference, uses the model
\begin{align}
y_i | \eta_i &\sim \m N(0, \sigma_\epsilon^2), \\
\eta_i &=  \beta_0 + u(s_i),
\end{align}
where $u(s)$ is the SGF.
For the simulation, $u(s)$ is the test surface that we aim to reconstruct, and $s_i$ are sampled at random.
For MS, MB, and MN, $u(s)$ is as described in Section 4, 
and the prior for the hyper-parameters follow the same rationale as for the application in Supplementary Section \ref{app-hyper-prior}.
The specific values chosen for the prior parameters are the same for all models,
$\lambda_\epsilon=\lambda_{\sigma_u} = 1.5$,
and we set the prior median of the spatial range to 1.

\subsection{The prior model for the Archipelago dataset}

\label{app-arch-prior-model} 

One of the advantages with the Bayesian framework is that it is straightforward to detail all the model assumptions, as they can all be written down as prior probability distributions.
In this section we detail the prior models used in Section 5 and their rationale.

\subsubsection{The predictor}

The prior models are,
\begin{align}
y_i | \eta_i &\sim \up{Poisson}(z_ie^{\eta_i}) \\
\label{eqFullModel}
\eta_i &=  \vec x_i \vec \beta + u(s_i) + \epsilon_i \\
\epsilon_i &\sim \m N(0, \sigma_\epsilon^2 )
\label{eq-bayesian-models}
\end{align}
where the observations are assumed to be conditionally independent given the linear predictor $\vec \eta$ (which corresponds to the log intensities) and the fixed sampling effort $z_i$. 
The vector $\vec x_i$ contains the covariate values, $\vec\beta$ is the vector of fixed effects (including intercept), $u(s)$ denotes the spatial random effect, and  $\epsilon_i$ is the iid random effect. 
The prior for $\beta$ is almost flat, implemented as a Gaussian with a very large variance.

Typically in ecological data the assumption of linear relationship between mean and variance in the observation process, implied by the Poisson distribution, is too restrictive \citep{Linden+Mäntyniemi:2011}. 
This extra variation (over-dispersion) over the Poisson distribution, arising from, e.g., aggregation behavior of individuals, and sampling variability, is modeled with the iid random effect $\epsilon_i$. 
The SGF $u(s)$ models spatial variability, such as diffusion and spatial aggregation of the species, effect of covariates that have not been accounted for, and spatially correlated measurement errors.

For the spatial effect $u$, we use the stationary SGF, the Barrier SGF, and a non-spatial model ($u=0$).
This is because the stationary model is the current standard choice in R-INLA,
and the non-spatial model ($u=0$) we use as reference model when comparing predictive performance.

\subsubsection{The priors for the hyper-parameters ($r, \sigma_u, \sigma_\epsilon$)}

\label{app-hyper-prior}

Here we provide a detailed justification for the priors on the hyper-parameters in the case of the smelt larvae. 
Similar reasoning also holds for the horseshoe example.

\label{sec:prirosOnHyperparameters}

We use the following priors for the hyper-parameters
\begin{align}
\pi (\sigma_\epsilon) &\sim \lambda_\epsilon e^{-\lambda_\epsilon \sigma_\epsilon} \\
\pi (\sigma_u) &\sim \lambda_0 e^{-\lambda_0 \sigma_u} \\
\pi \left(\frac{1}{r}\right) &\sim \lambda_1 e^{-\lambda_1 \frac{1}{r}},
\label{priors}
\end{align}
where the choice of $\lambda$s depends on the problem's scale and the user's desire for shrinkage. 
In this section we discuss the reasoning behind choosing these priors, and define $\lambda$s for our case study.

We consider it appropriate to set priors based on \emph{prior modeling preferences}. With this we do not mean that you can elicit the prior completely from \emph{a priori} information about the parameter, but by \emph{a priori} information about how you want the model to behave.
As is common in applied models, SDMs are not looking at one specific parameter of interest, but rather several different parameters and a diverse range of possible prediction scenarios.
Our focus then is to avoid over-fitting, and to ensure that simple models are recovered when they are true (i.e.\ we are more concerned about estimating the parameters correctly when the true value represents a simpler model).
First, we will discuss the chosen priors as shrinkage priors, 
then we will consider if this is consistent with the interpretation of the hyper-parameters, 
and we will set the $\lambda$ parameters.

To construct shrinkage priors, we need 
\begin{enumerate}
\item the concept of a base model, that is, a simple model to shrink towards,
\item an understanding of what models are essentially the same, and what models are different, so that one can have a meaningful way of choosing the shape of the shrinkage prior, and,
\item a degree of shrinkage that may vary between applications (essential for taking into account the scale at which data are measured, and the number of competing model components).
\end{enumerate}

A good way to achieve these goals is found in \citet{Simpson2014penalising}, 
which is built upon by \citet{Fuglstad2015interpretable} to give priors for range and standard deviation of a Mat\' ern field. 
Using their results, we get the priors in equation \eqref{priors}, with the following details.
\begin{enumerate}
\item The base model is $\sigma_\epsilon=0, \sigma_u = 0$, and $r = \infty$.
\item Distance from the base model, as measured by the square root of the Kullback-Leibler divergence, is a parameter where an exponential prior is  reasonable .
\item The strength of shrinkage is chosen by setting $\lambda$'s.
\end{enumerate}

Now for the question of whether these priors are in line with the interpretation of our model.
When including an additive component in the linear predictor, one must have an understanding of what structure it is supposed to model.
This understanding naturally leads us to prefer some model components for modeling the structure of the data. 
In general, we strongly prefer modeling variation (in log intensity) with the fixed effects over using random effects, which leads to shrinkage priors for the variance parameters $\sigma_u$, $\sigma_\epsilon$. 
The strength of this preference is tuned by $\lambda$
(e.g.\ small values of $\lambda$ lead to greater use of random effects, which possibly leads to weaker conclusions about the fixed effects).

For the size of a model component, we believe that $\sigma$ is a very good and interpretable parametrization since it has the same units as the linear predictor, and since $\sigma \approx 0$ represents the removal of the model component.
The exponential distribution is a reasonable prior due to its memoryless property,
which represents the idea that additional randomness should be penalized in the same way, independent of how much randomness you have already included in the model.

The two random effects, the iid effect $\epsilon$ and the spatial effect $u$, are both representing noise in observations, noise in covariate measurements, species aggregation, and missing covariates, respectively, in a local or spatially correlated manner. 
\emph{A priori} we do not know whether the local or spatially correlated effect should dominate; this 
leads us to having no preference between the two model components, and so $\lambda_\epsilon=\lambda_{\sigma_u}$. 
To determine the value of $\lambda_\epsilon$ we look at the maximum size of the iid random effect.
We believe that the iid effect should be able to explain any datapoint on its own, but that it should be \emph{a priori} extremely difficult, say with a probability of $0.01$.
The maximum logarithmic value of our measurement counts is from around 4 to around 5 in the three datasets. 
Because we have an intercept in the model, we think the random effect should only need to extend up, or down, 3 logarithmic units (we use the natural logarithm) in any of the datasets.
We compute the marginal distribution
$$\pi(\epsilon_i) = \int \pi(\epsilon_i | \sigma ) \pi(\sigma )\d \sigma $$
numerically and define $\lambda_\epsilon = 1.5$ to get the 1\% and 99\% quantiles approximately at -3 and 3.

This choice is obviously somewhat arbitrary, and different ways of reasoning will give different values.
We found values from 0.9 all the way to 2 to be reasonable.
However, the results are not unduly influenced by this, since the models with $\lambda =0.9$ or $\lambda=2$ give reasonably close parameter estimates to that of the chosen model ($\lambda_\epsilon=1.5$).
Further, this paper's main focus is on comparing the stationary and the barrier model, where we use the same priors in both models.
The interpretation of this prior, for the Barrier model with a non-stationary marginal variance, is slightly different from the stationary model; in the Barrier model, the penalty is not constant over space.
However, the spatial variability of the penalty is less important than the variability between different reasonable choices of $\lambda$.

For the range $r$ of the spatial effect we have a prior preference for longer range.
That is because, by definition, the spatial effect is there to pick up non-local variability. 
In truth, we expect there to be variability at several different spatial scales, but the spatial effect is supposed to pick up the long range dependencies, leaving the very short range dependencies to the iid effect.
However, we want the prior to be weak, so that it is dominated by the likelihood when the short scale spatial structure is the most important for explaining the observations.
From a practical point of view, the parametrization $1/r$ can seem strange, and the $r$ parameter can seem more natural.
The problem with the $r$ parameter is that the model $r=1$ and the model $r=2$ seem very close, but they are very different models, with very different behaviour.
And although $r=10$ and $r=20$ seem very far apart ($\Delta r = 10$), the difference between the behaviour at these values are less than the difference between the behaviour at $r=1$ and $r=2$.
The $1/r$ parametrization solves these issues neatly.
In this parametrization it is also possible to see when the posterior is leaning towards the base model ($\frac{1}{r} \approx 0$).

For setting $\lambda_1$ we first need to set a value for the typical length of the study area.
Then we claim that half of this typical length is a decent choice for the prior median of the range, 
because when sampling spatial fields with this range, the fields seem to be reasonable models of long range spatially structured ecological effects.
In the case study, the typical length is 60km, giving prior median range 30km, and $\lambda_1 = 20.8 \up{km}$. 
Additionally, we have a consideration about the tail behaviour of the prior.
Around $r=4\up{km}$ we consider the spatial model to have failed to produce reasonable results, because the mesh discretization does not adequately resolve the structure at this scale (this can be fixed by using a finer mesh).

We leave the topic of priors with a few additional notes.
We have chosen to use the same prior in the stationary model and in the barrier model, but it would be possible to use a slightly larger $\lambda_1$ in the barrier model, estimating a slightly longer range.
We also recommend these priors, or the equivalent likelihood penalties, as in our experience they make the posteriors better behaved, and stabilize the inference in applications with Poisson observation models.

\subsection{Pike and pikeperch in the Archipelago}

\label{app-allmodels}

In the main part of the paper, we only include selected results for the data on the smelt species; here we present the other results as well.

\subsubsection{Analysis of the perch data}
\begin{figure}
\centering
\includegraphics[trim=7mm 7mm 0mm 0mm,clip,width=1\linewidth]
{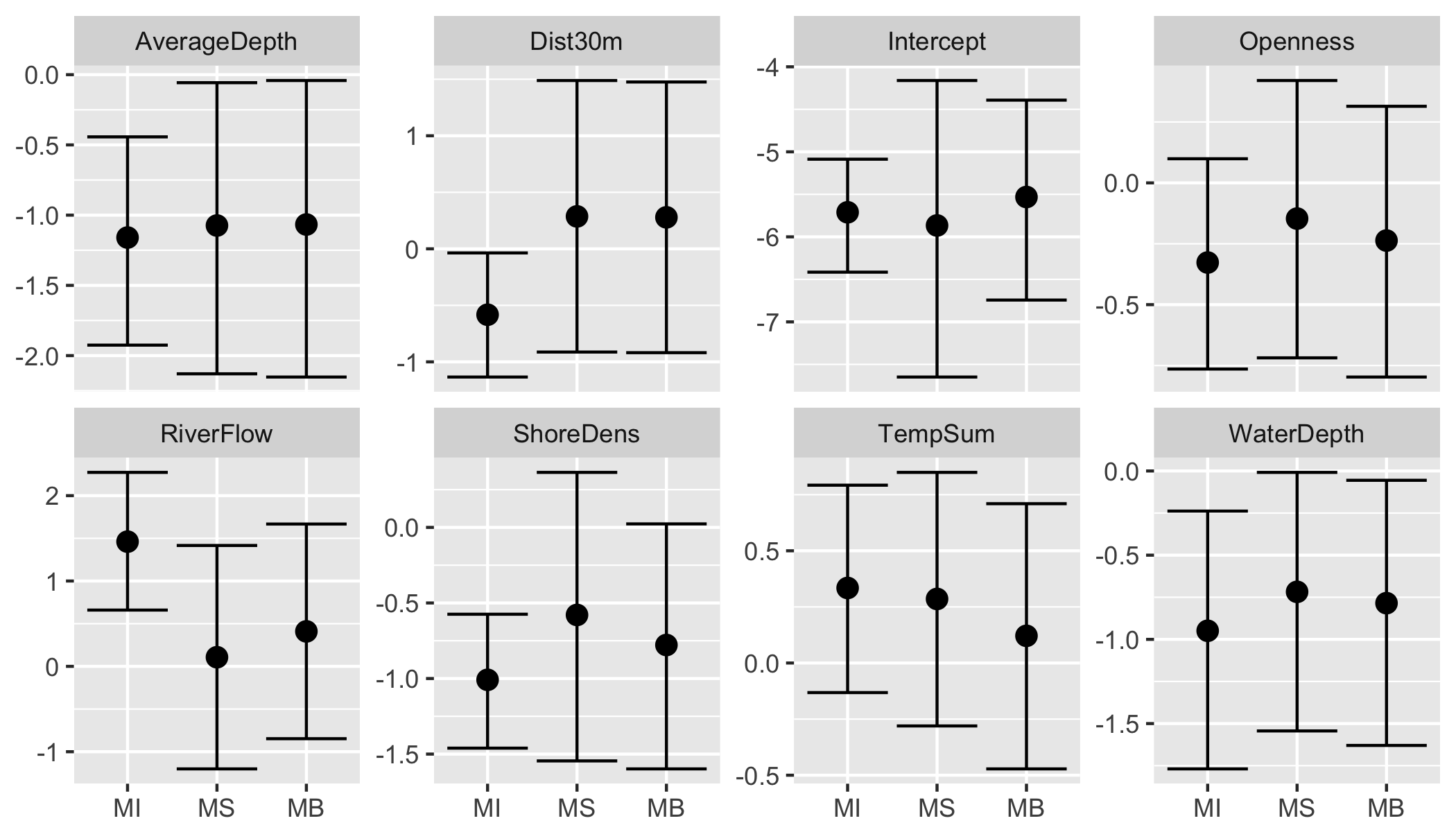}
\caption{
The posterior medians and 95\% credible intervals of the fixed effects in the perch analysis between the three models. 
$\m M_I$ denotes the model without spatial field, 
$\m M_S$ denotes the model with stationary spatial field, and $\m M_B$ denotes the model with (non-stationary) spatial field with barriers.
}
\label{fig-covars-perch}
\end{figure}

The estimates of the fixed effects have a similar structure for perch, see Figure \ref{fig-covars-perch}, as it did for smelt. 
The magnitude of the estimated impact of each covariate is less, 
but the significance levels of the covariates are similar.

\begin{figure}
\centering
\begin{subfigure}[b]{0.7\textwidth}
\includegraphics[width=\linewidth]
{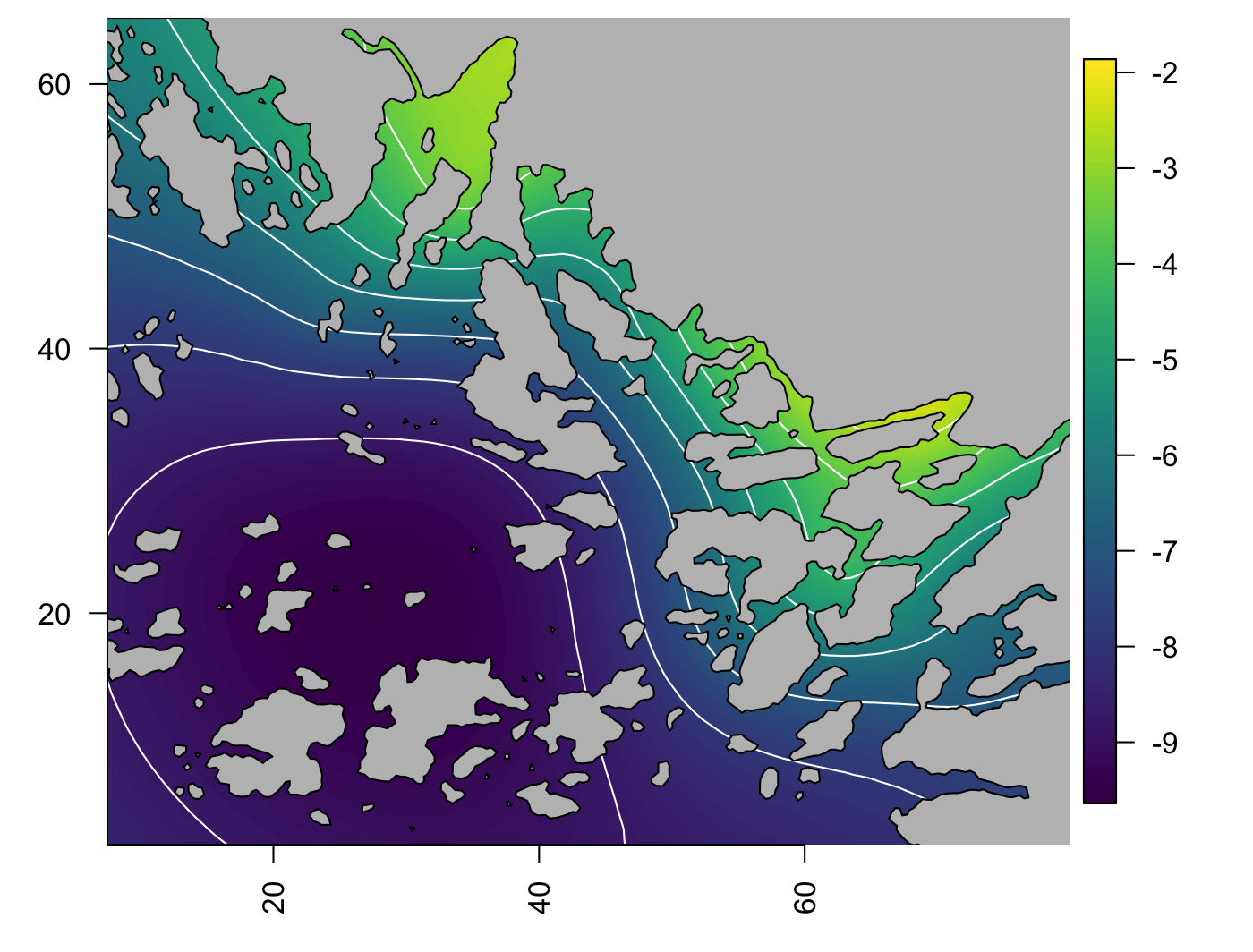}
\caption{Stationary model, posterior mean}
\label{fig:perch-field01}
\end{subfigure}

\begin{subfigure}[b]{0.7\textwidth}
\includegraphics[width=\linewidth]
{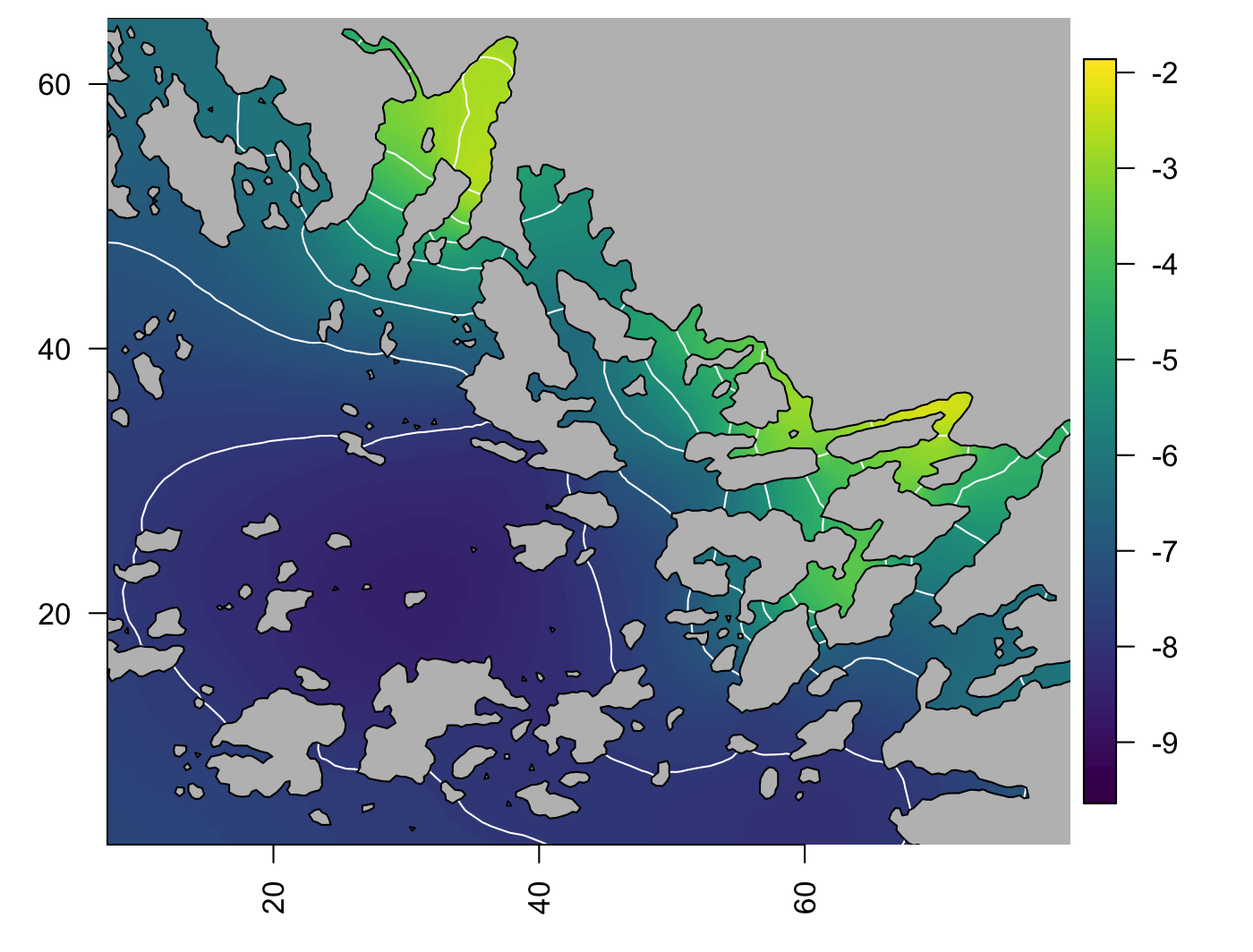}
\caption{Barrier model, posterior mean}
\label{fig:perch-field02}
\end{subfigure}

\caption{Posterior estimates of the spatial field for the perch larvae in the models without covariates.
We see how the stationary model oversmooths in the northern part of the figure compared to the barrier model. 
}
\label{fig:perch-ResultSpaceNocov-1}
\end{figure}

\begin{figure}
\centering
\begin{subfigure}[b]{0.7\textwidth}
\includegraphics[width=\linewidth]
{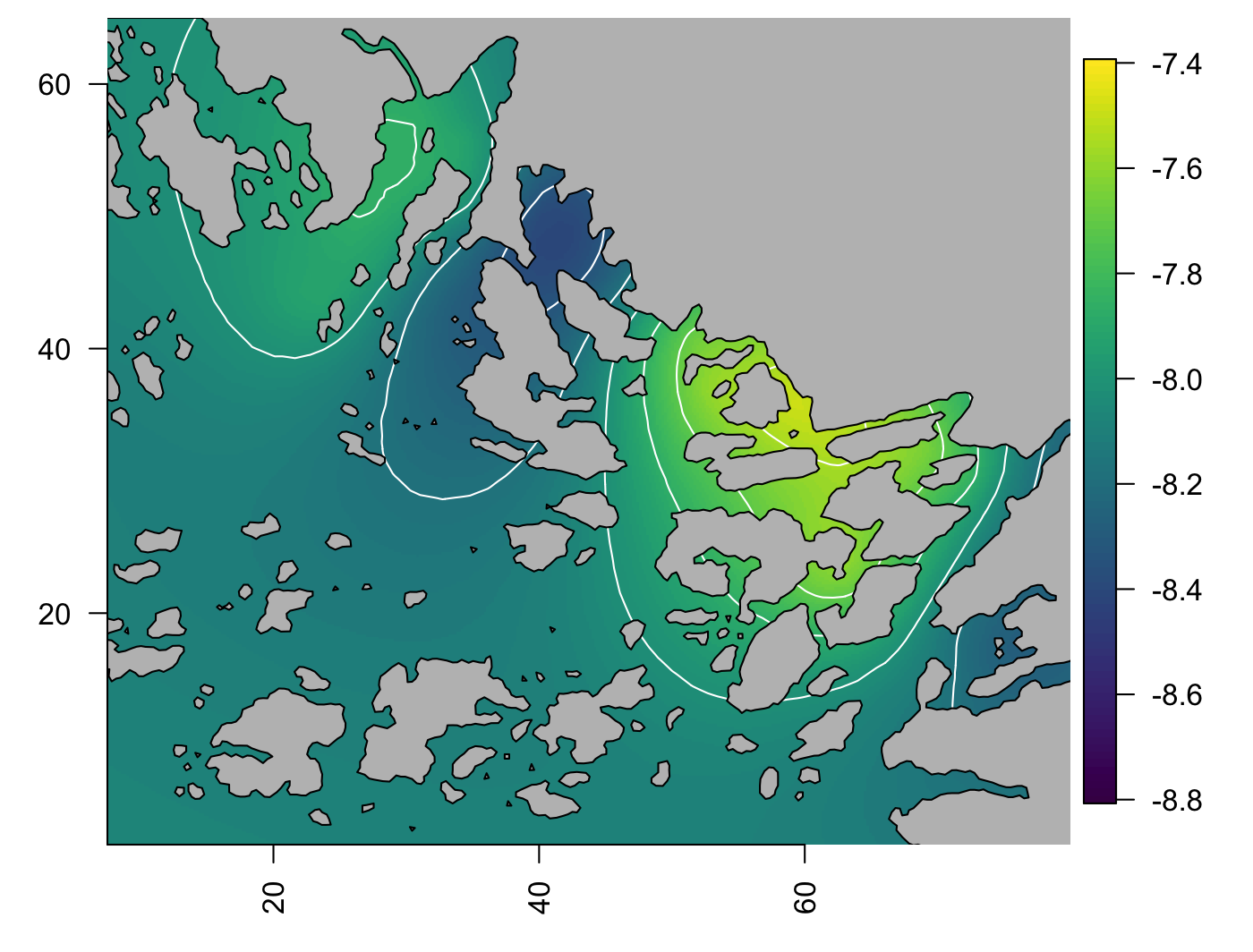}
\caption{Stationary model, posterior mean}
\label{fig:perch-field11}
\end{subfigure}

\begin{subfigure}[b]{0.7\textwidth}
\includegraphics[width=\linewidth]
{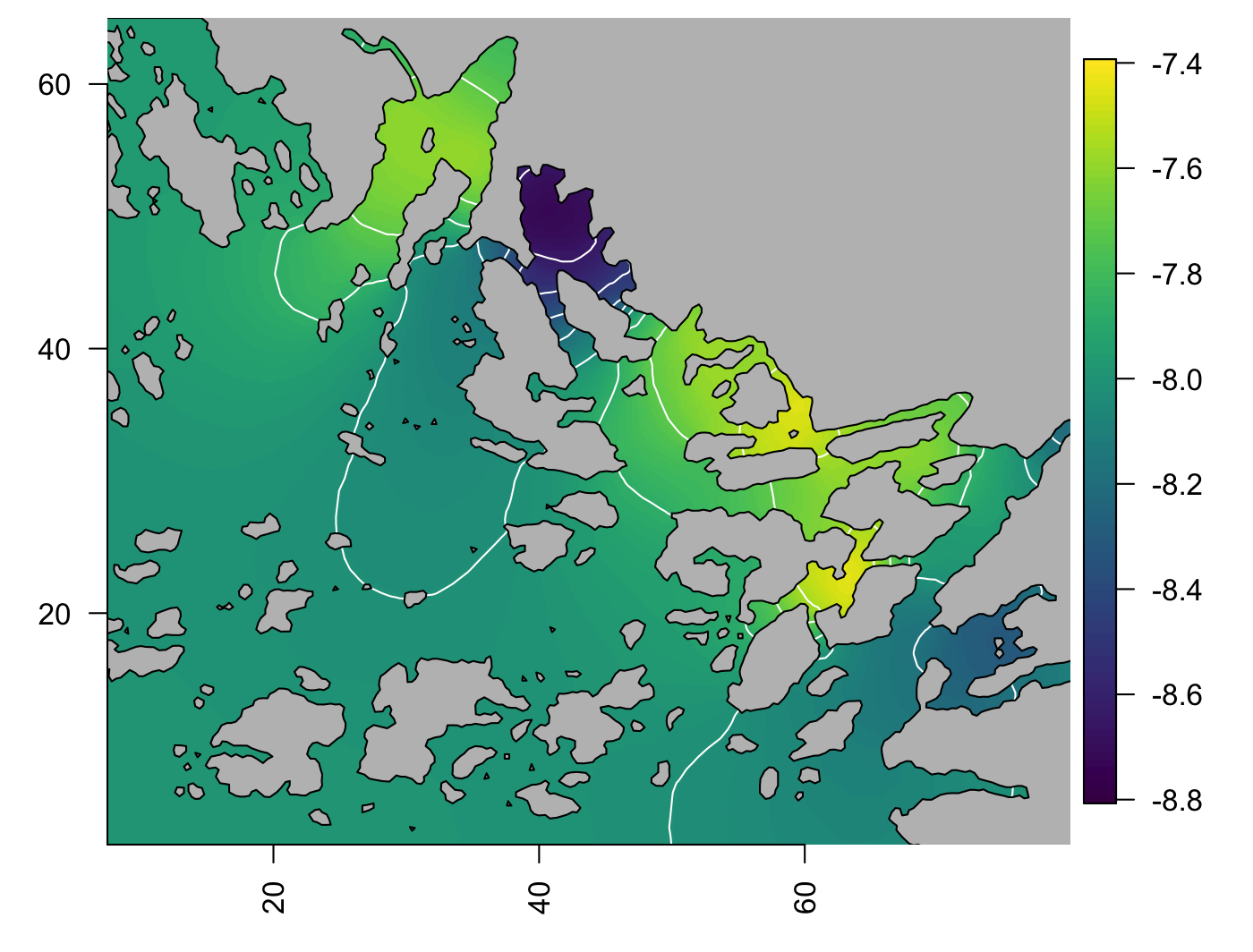}
\caption{Barrier model, posterior mean}
\label{fig:perch-field12}
\end{subfigure}

\caption{Posterior estimates of the spatial field for the perch larvae in the models with covariates.
We see how the stationary model oversmooths in the northern part of the figure compared to the barrier model. 
}
\label{fig:perch-ResultSpaceWithcov-1}
\end{figure}

In the spatial plots in figures \ref{fig:perch-ResultSpaceNocov-1} and \ref{fig:perch-ResultSpaceWithcov-1},
we see that the perch larvae has a similar species distribution to the smelt larvae, but with an additional hotspot in the south-east.

The level of similarity between the two datasets hints that the species are co-occurring.
The changes in fixed effects from MI to MS to MB in the perch case is also very similar to the corresponding changes for smelt.
This hints that the bias (or bias correction) in the fixed effects may be driven by a confounding between the spatial effect and the covariates, or by the same missing covariates.

\subsubsection{Analysis of the pikeperch data}
\begin{figure}
\centering
\includegraphics[trim=7mm 7mm 0mm 0mm,clip,width=1\linewidth]
{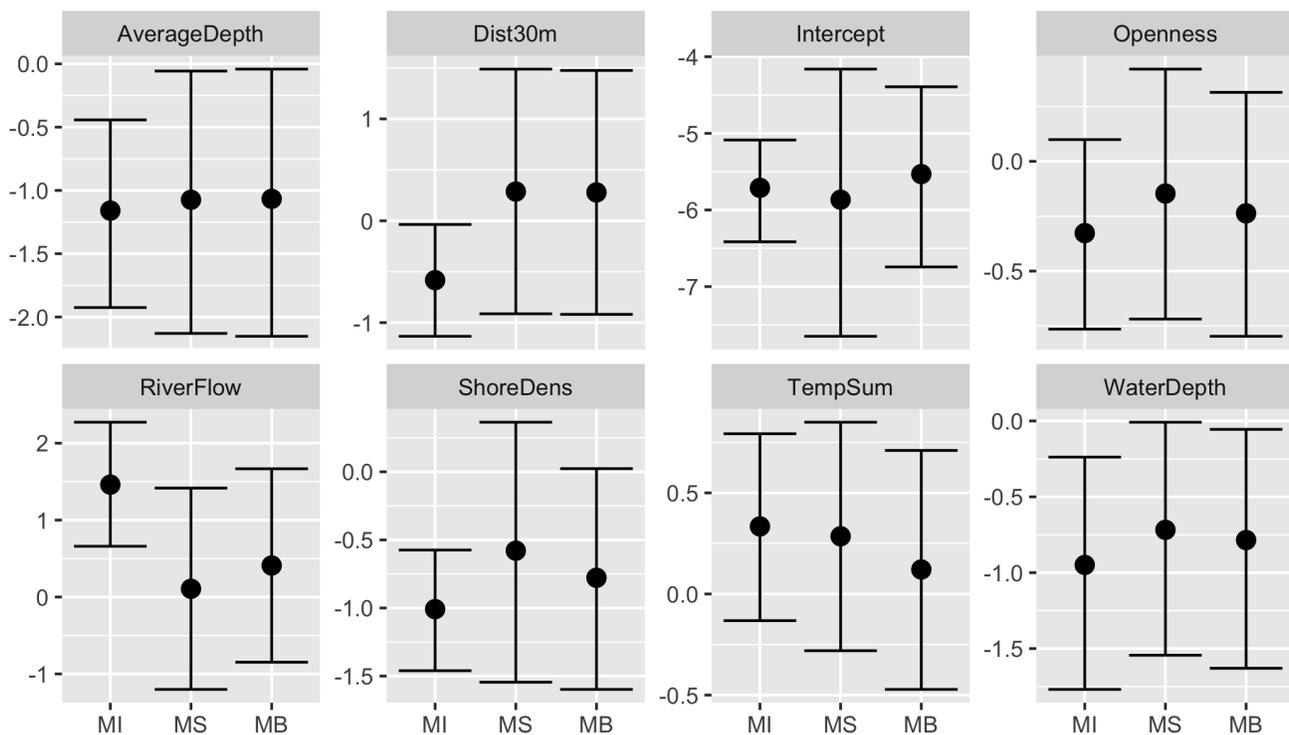}
\caption{
The posterior medians and 95\% credible intervals of the fixed effects in the pikeperch analysis between the three models. 
MI denotes the model without spatial field, 
MS denotes the model with stationary spatial field, 
and MB denotes the model with (non-stationary) spatial field with barriers.
}
\label{fig-covars-pikep}
\end{figure}

The posterior fixed effects for the pikeperch larvae are in Figure \ref{fig-covars-pikep}.
These results are different than for the smelt larvae for the two covariates RiverFlow and ShoreDens.
The significance of these two covariates are a lot less in the pikeperch analysis.
We also see larger differences between stationary and Barrier model for the pikeperch data than what we see for the smelt data.

\begin{figure}
\centering
\begin{subfigure}[b]{0.7\textwidth}
\includegraphics[width=\linewidth]
{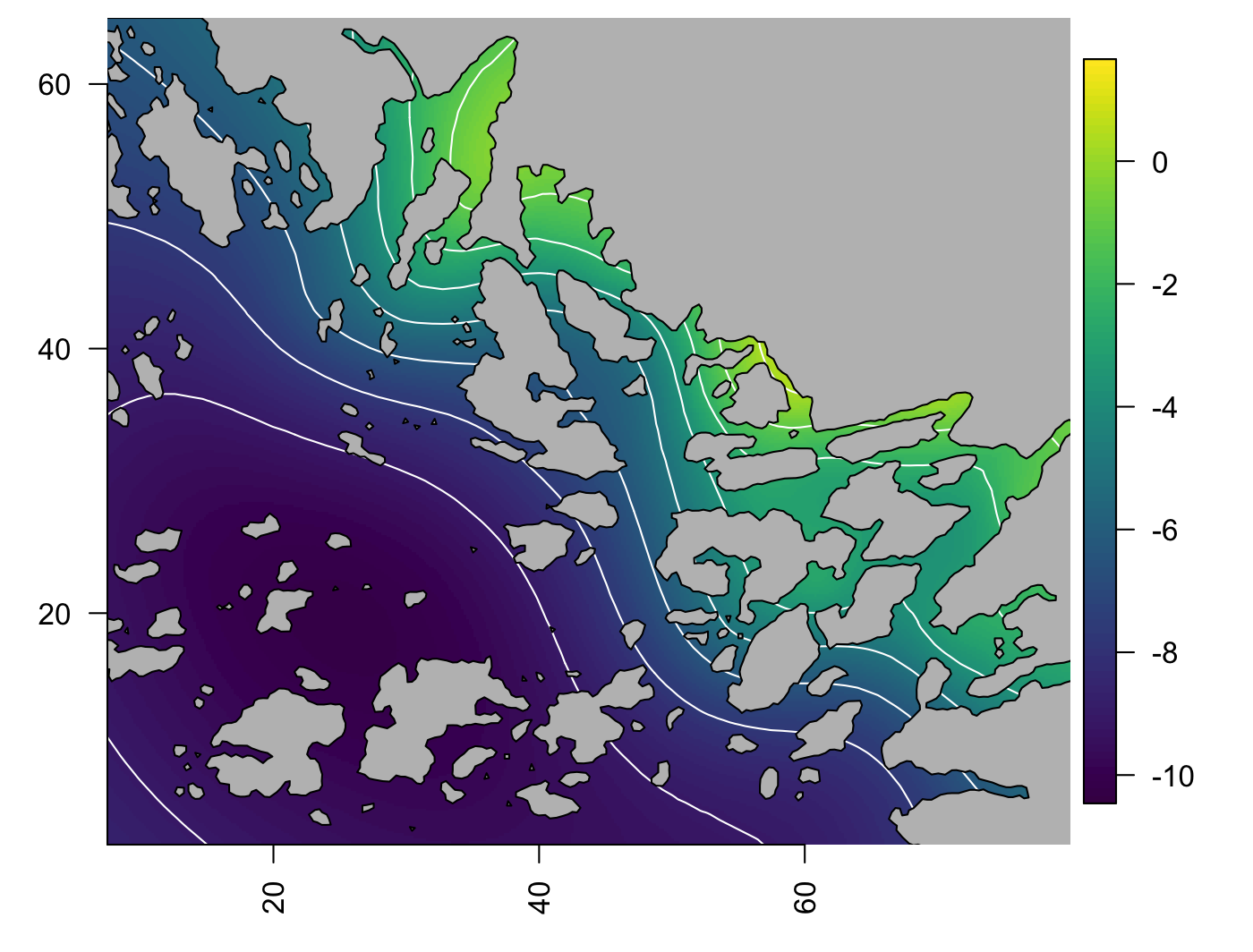}
\caption{Stationary model, posterior mean}
\label{fig:pikep-field01}
\end{subfigure}

\begin{subfigure}[b]{0.7\textwidth}
\includegraphics[width=\linewidth]
{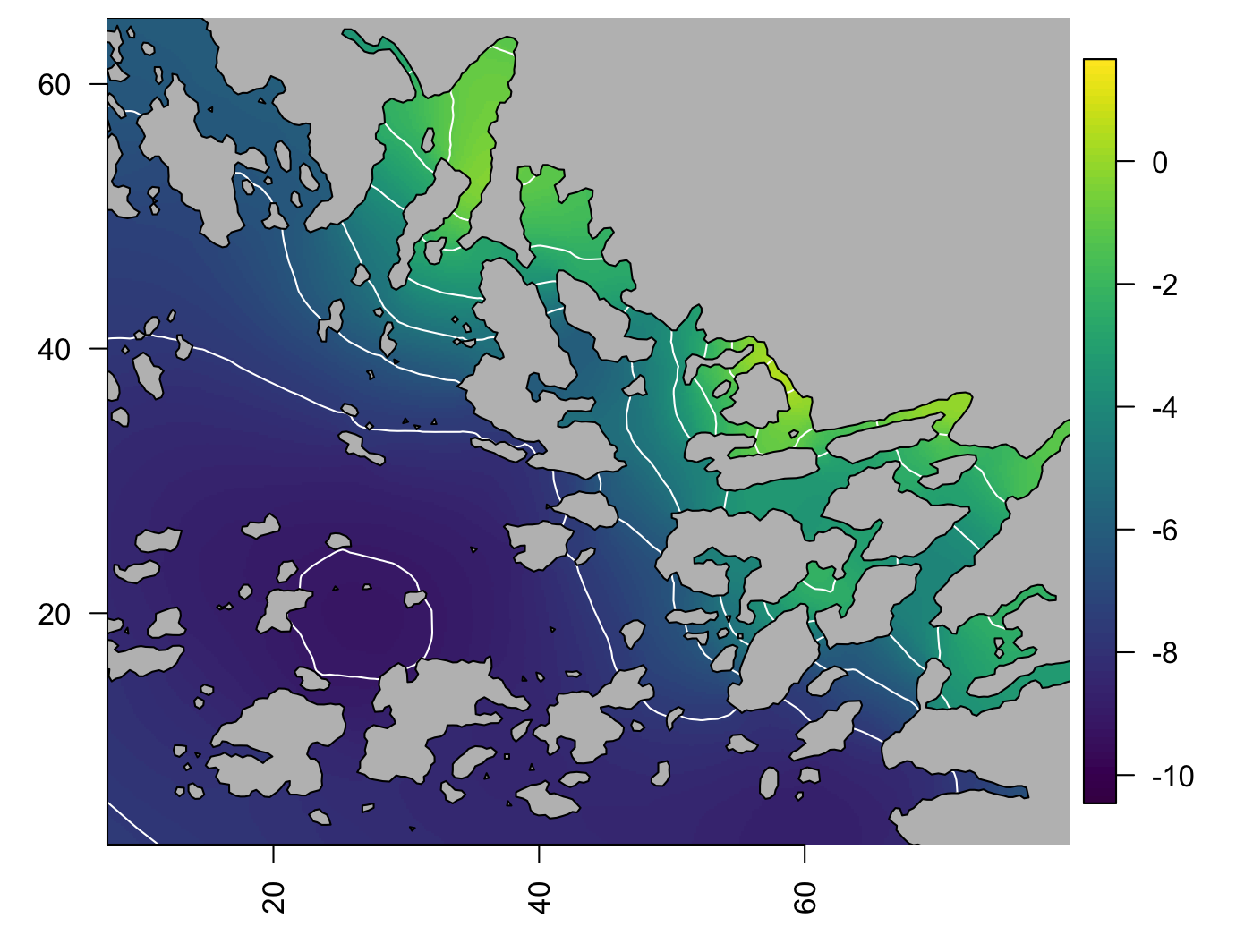}
\caption{Barrier model, posterior mean}
\label{fig:pikep-field02}
\end{subfigure}
\caption{Posterior estimates of the spatial field for the pikeperch larvae in the models without covariates.
There are only minor differences in the estimates. 
}
\label{fig:pikep-ResultSpaceNocov-1}
\end{figure}

\begin{figure}
\centering
\begin{subfigure}[b]{0.7\textwidth}
\includegraphics[width=\linewidth]
{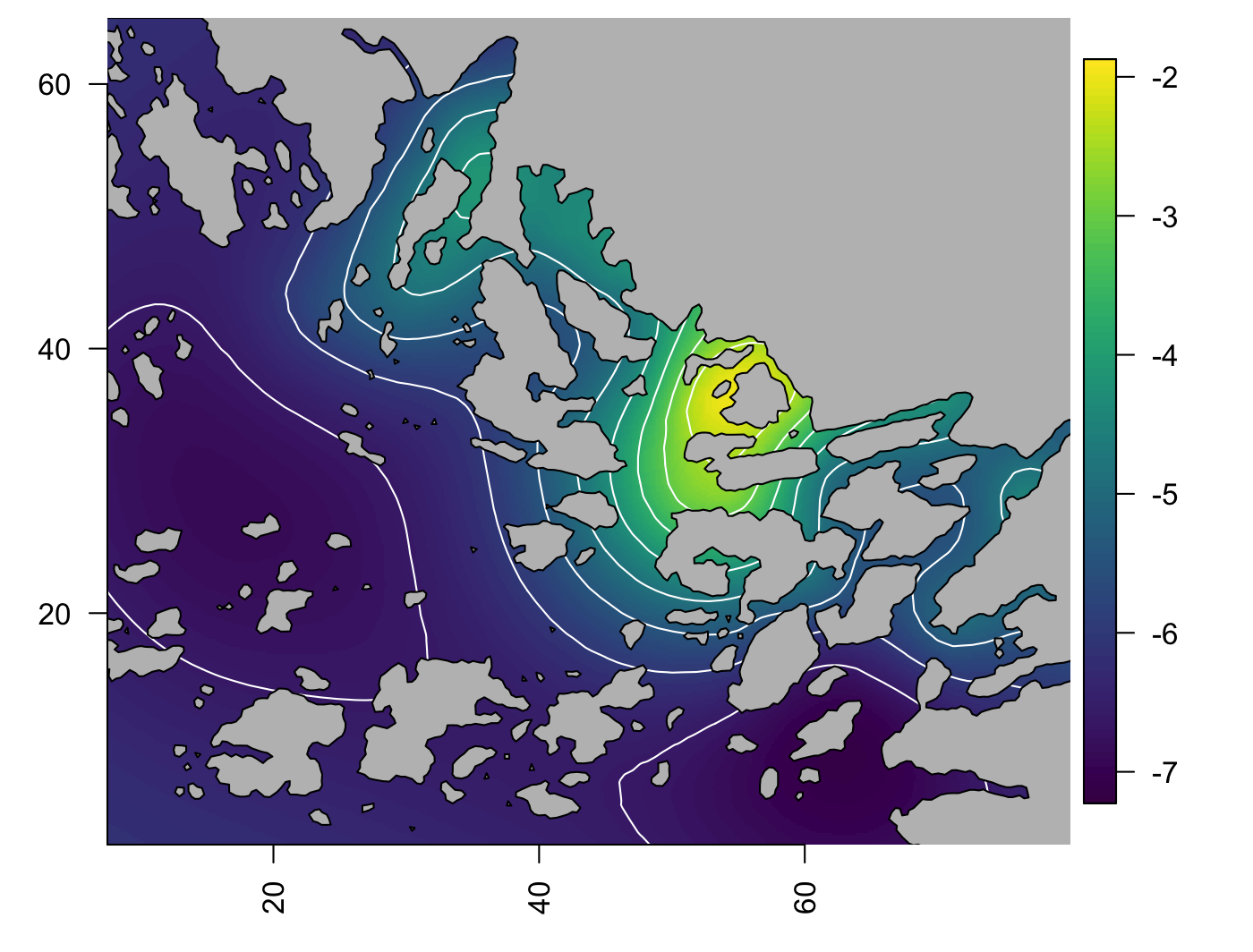}
\caption{Stationary model, posterior mean}
\label{fig:pikep-field11}
\end{subfigure}

\begin{subfigure}[b]{0.7\textwidth}
\includegraphics[width=\linewidth]
{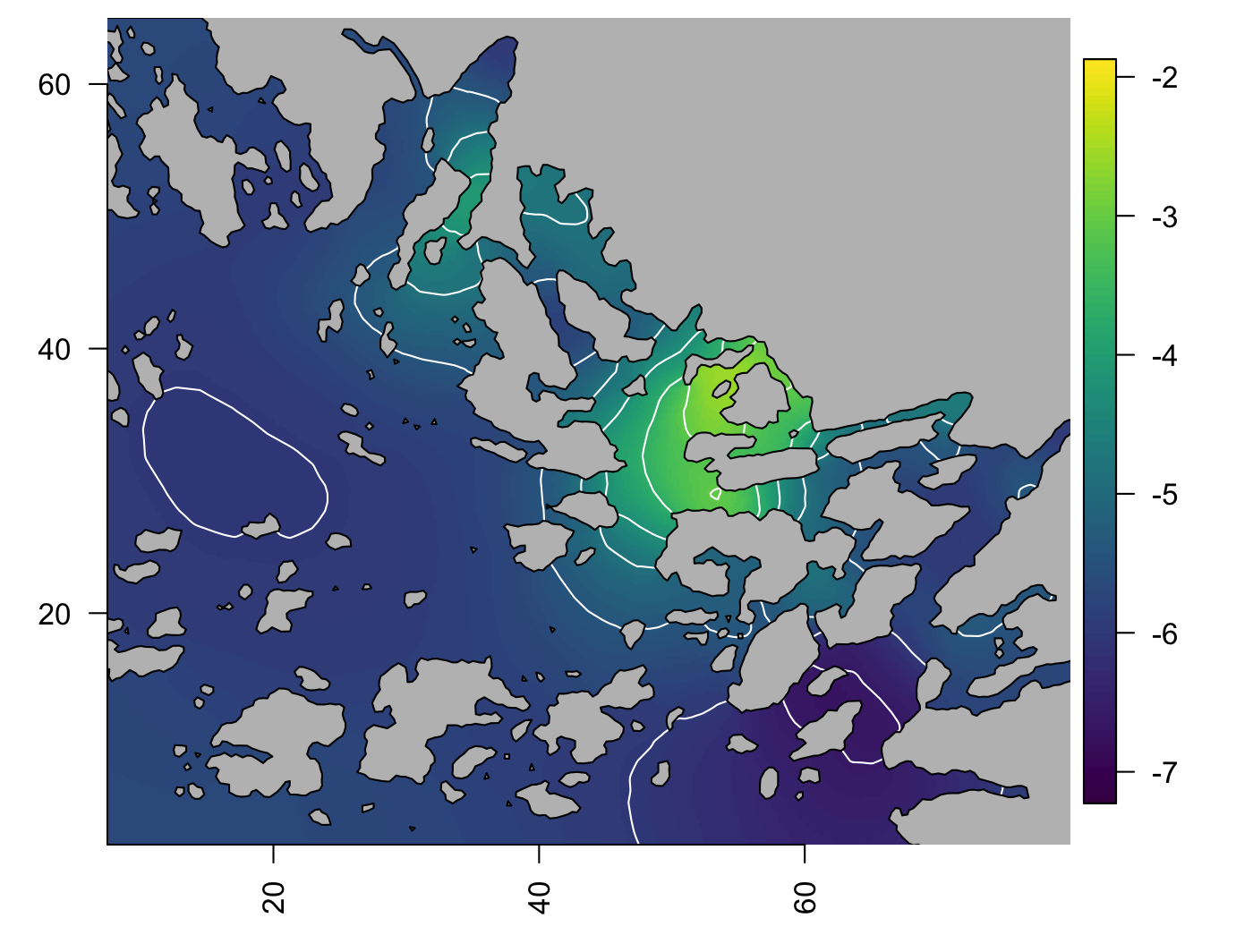}
\caption{Barrier model, posterior mean}
\label{fig:pikep-field12}
\end{subfigure}

\caption{Posterior estimate (median) of the spatial field for the perch larvae in the models with covariates.
The differences can be seen in the north-west and the east of the map.
}
\label{fig:pikep-ResultSpaceWithCov-1}
\end{figure}

The posterior spatial effects are visualised in figures \ref{fig:pikep-ResultSpaceNocov-1},  
\ref{fig:pikep-ResultSpaceWithCov-1} and.
For the pikeperch, there is no longer any oversmoothing between the two inlets in the northern part of the map. 
We see that the abundance happens to be large on both sides of the land barrier, and therefore that the two models produce very similar estimates in the northern region.

\subsection{Model comparison for the archipelago}

\label{app-arch-loocv}

In this appendix we show the full details of the model comparison that was described at the end of section 5.
LOOCV means Leave-One-Out Cross Validation; holding out one datapoint and fitting the model to the rest, then using the inference to predict the held-out point.
The computations here are not approximations; we ran 3564 models.
NLPD means Negative Log Predictive Density and is how the predictive performance is measured for each LOOCV point, i.e.
$$\text{NLPD}_i = -\log \pi(y_i | y_{-i}). $$
The models are named M1 through M6 as follows.
M1 is the stationary model without covariates, M2 is the stationary model with covariates, M3 is the barrier model without covariates, M4 is the barrier model with covariates, M5 is the IID model without covariates, and M6 is the IID model with covariates.

\begin{figure}
\centering
\includegraphics[trim=7mm 7mm 0mm 0mm,clip,width=1\linewidth]
{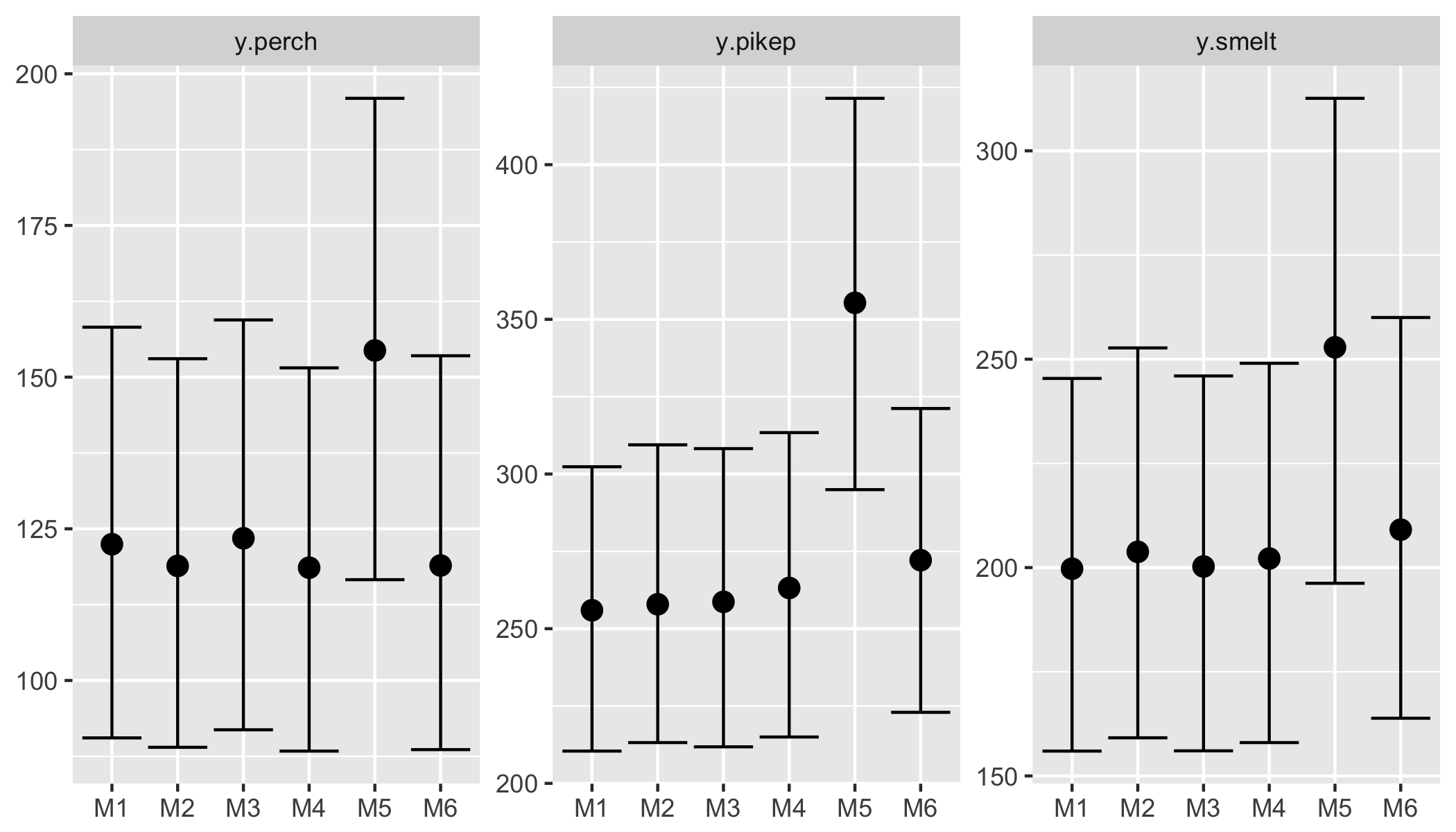}
\caption{Bootstrapped mean LOOCV NLPD for the three datasets.
See appendix \ref{app-arch-loocv} for details on the models M1 to M6.
}
\label{fig-loocv-bootstrap}
\end{figure}

\begin{figure}
\centering
\includegraphics[trim=7mm 7mm 0mm 0mm,clip,width=1\linewidth]
{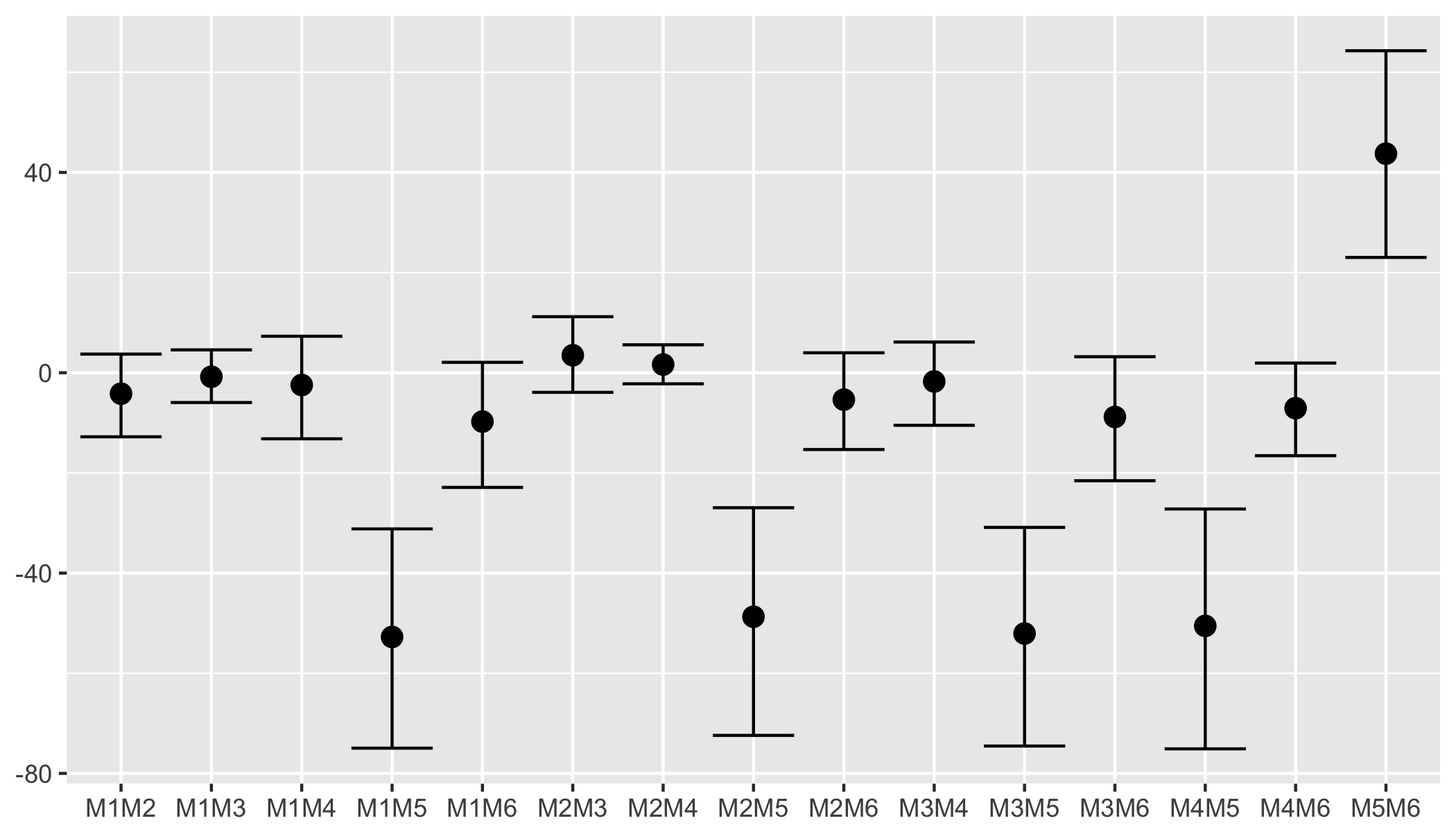}
\caption{Bootstrapped mean differences in LOOCV NLPD for the smelt data.
See appendix \ref{app-arch-loocv} for details on the labels.
}
\label{fig-loocv-bootstrap-diff-smelt}
\end{figure}

\begin{figure}
\centering
\includegraphics[trim=7mm 7mm 0mm 0mm,clip,width=1\linewidth]
{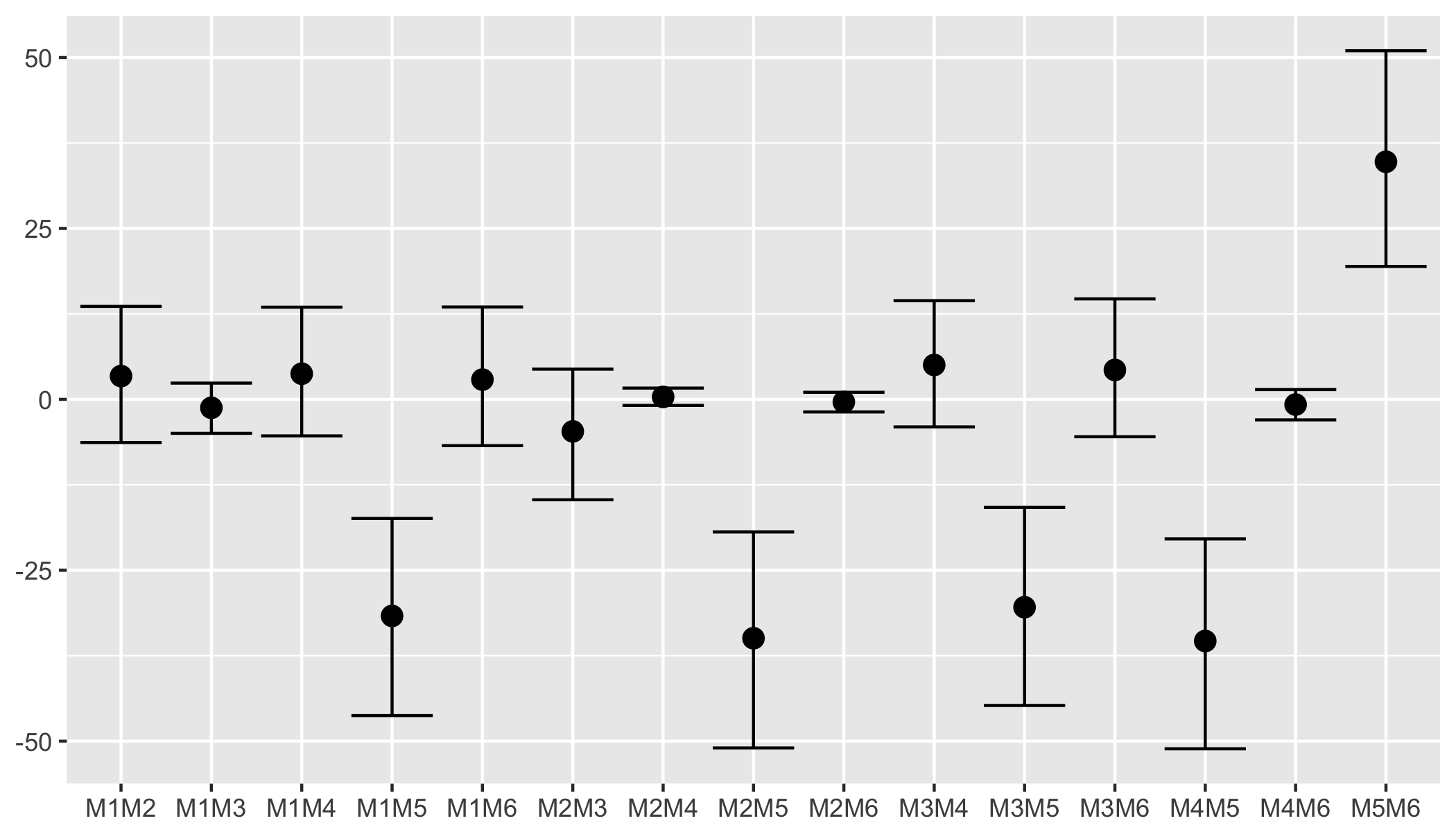}
\caption{Bootstrapped mean differences in LOOCV NLPD for the perch data.
See appendix \ref{app-arch-loocv} for details on the labels.
}
\label{fig-loocv-bootstrap-diff-perch}
\end{figure}

\begin{figure}
\centering
\includegraphics[trim=7mm 7mm 0mm 0mm,clip,width=1\linewidth]
{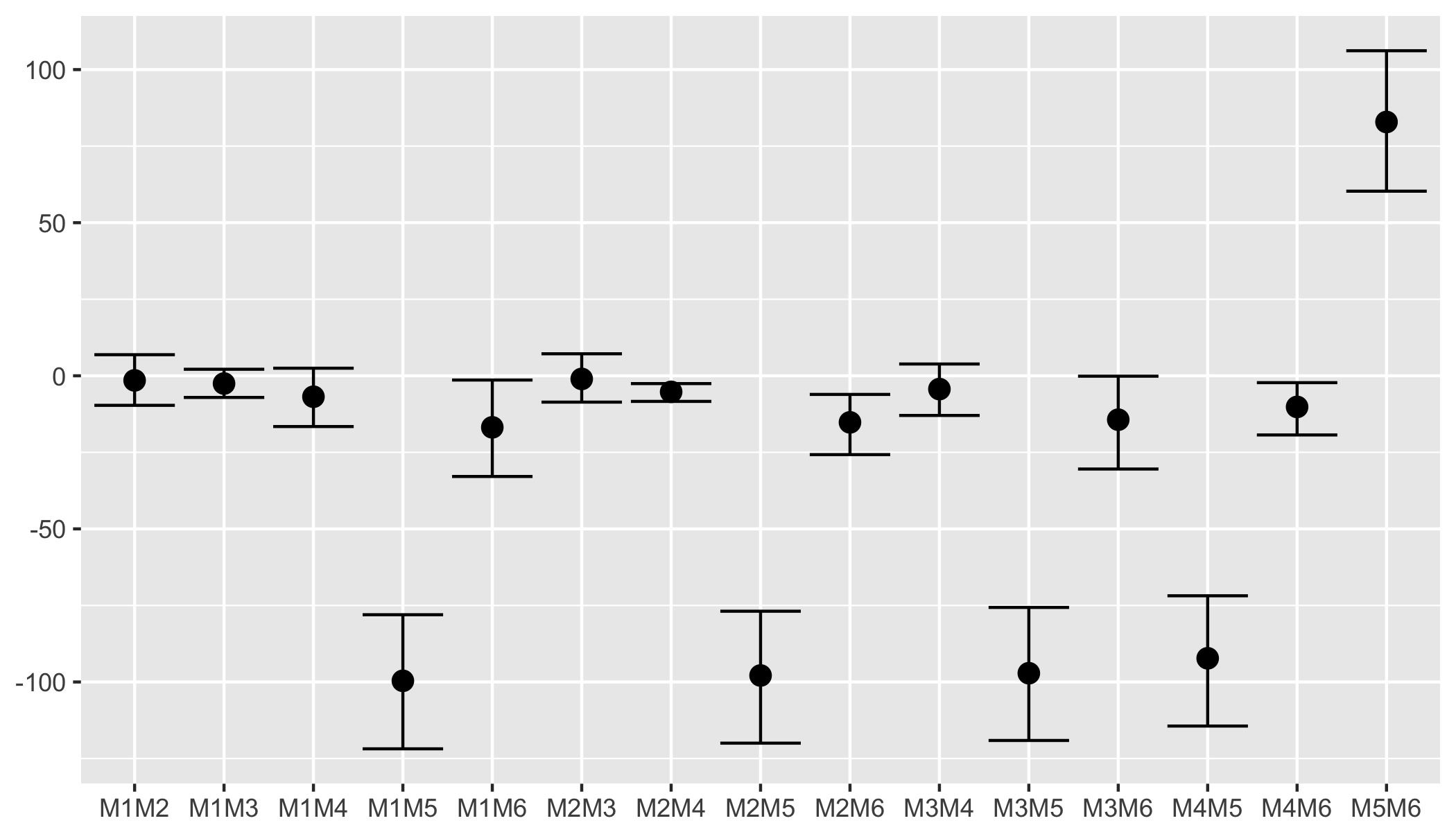}
\caption{Bootstrapped mean differences in LOOCV NLPD for the pikeperch data.
See appendix \ref{app-arch-loocv} for details on the labels.
}
\label{fig-loocv-bootstrap-diff-pikeperch}
\end{figure}

Figure \ref{fig-loocv-bootstrap} shows the individual scores, with bootstrapped median and 95\% equal tailed intervals.
Figures \ref{fig-loocv-bootstrap-diff-smelt}, \ref{fig-loocv-bootstrap-diff-perch} and \ref{fig-loocv-bootstrap-diff-pikeperch} shows all the bootstrap medians and intervals for the pairwise differences, where e.g.\ M1M4 is the difference between M1 and M4.

In all the difference figures, we observe that M5 is a statistically significant bad model.
This is not surprising as this is a model only with intercept and individual errors.
The other significant results are found in the pikeperch models; figure \ref{fig-loocv-bootstrap-diff-pikeperch}. 
For the pikeperch data, all the spatial models are better than the non-spatial models.
In total, we are computing 30 confidence intervals (when excluding the M5 models), at 95\%, hence, we expect a few of the intervals to randomly exclude 0 (and give false negatives).
From looking at the plots we conclude that none of the significant results are clear enough to be considered strong evidence of one model outperforming another.


\subsection{Simulation study in the archipelago}
\begin{figure}
	\centering
	\begin{subfigure}[b]{0.5\textwidth}
		\includegraphics[width=\linewidth]
		{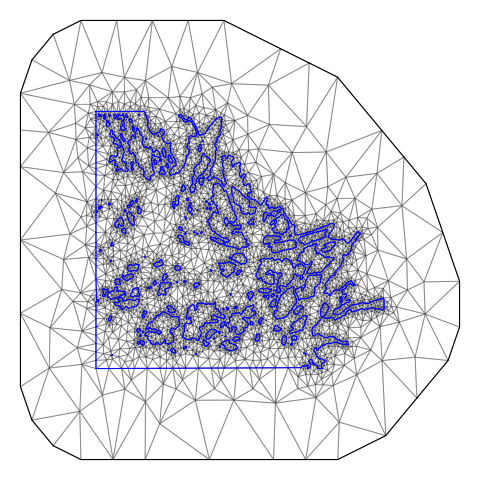}
		\caption{Mesh used for simulation}
		\label{fig:archipsim-1}
	\end{subfigure}
	
	\begin{subfigure}[b]{0.5\textwidth}
		\includegraphics[width=\linewidth]
		{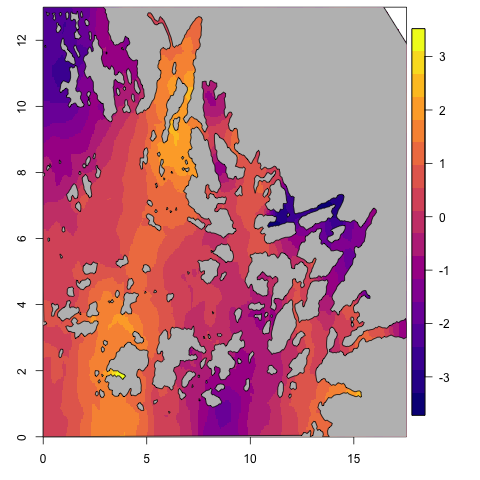}
		\caption{Simulated ``true'' spatial field.}
		\label{fig:archipsim-2}
	\end{subfigure}
	
	\caption{The mesh and simulation used in the simulation study.
	}
	\label{fig:archipsim}
\end{figure}

In this appendix we perform a simulation study to investigate the behaviour of the Barrier model when different meshes and barrier ranges ($r_b$) are used, and compare this to the results form the stationary model.
The simulated spatial field is shown in Figure \ref{fig:archipsim}, together with the mesh used in the simulation.
In addition to this mesh (mesh1), we create two coarser meshes (mesh2, mesh3) with different settings, leading to different representations of the coastline.
We sample 10000 locations at random, and simulate data $y$ by adding Gaussian noise (with sd 0.1) to the spatial field.

Four different models (for each mesh) are fitted to the data, the stationary model, and three Barrier models.
The three Barrier models use different $r_b/r$ (barrier range divided by range in water), with values 0.1, 0.2, and 0.3.
To compare the 12 variants we perform 2-fold cross-validation, where half the data are held out, and we compare point predictions to the ``true'' sampled data at the held out locations.
For model comparison criterion we use the root mean square error (RMSE).
This criterion is used for its simplicity, but it is not a proper scoring rule in the 5000 dimensional joint prediction space.

We show the results in Table \ref{tab-sim-arc}. 
All the Barrier models significantly outperform all the stationary models.
The Barrier models have very similar predictive accuracy, and we conclude that the predictive performance is robust to the choice of $r_b/r$ and to the mesh.
We note that models with a coarser mesh performs less well than with a fine mesh; the differences are not large compared to the uncertainty, but they are consistent across all models.

\begin{table}
	\centering

	\begin{tabular}{|c|c|c|c|}
		\hline
		   RMSE  $\cdot$ 1E3    & Mesh 1 & Mesh 2 & Mesh 3 \\ \hline
		Stationary  & 54.6 (1.6) & 56.1 (1.4) & 56.6 (1.3) \\ \hline
		Barrier 0.1 & 42.0 (0.8) & 44.4 (0.9) & 45.4 (0.8) \\ \hline
		Barrier 0.2 & 42.2 (0.9) & 44.7 (1.1) & 45.7 (1.0) \\ \hline
		Barrier 0.3 & 42.9 (1.1) & 45.4 (1.2) & 46.5 (1.2) \\ \hline
	\end{tabular} 
\caption{Model comparison on a simulated dataset in the archipelago with 2-fold cross-validation. 
	The values are the median of the repeated RMSE between the mean prediction and the held out observations.
	Barrier 0.1 denotes the Barrier model with $r_b/r=0.1$, and so on.
The values in parenthesis are the differences between the 0.75 quantile and the 0.25 quantile of the RMSE scores (the interquartile range).
For readability, the numbers have been multiplied by 1000.
}
\label{tab-sim-arc}
\end{table}

\bibliography{biblio-hb-18-01-09-plus}